\documentclass{aa}
\usepackage{natbib,psfig,graphicx,ulem}
\usepackage{txfonts}
\def\mathnew{\mathsurround=0pt}
\def\simov#1#2{\lower .5pt\vbox{\baselineskip0pt \lineskip-.5pt
\ialign{$\mathnew#1\hfil##\hfil$\crcr#2\crcr\sim\crcr}}}

\def\MeV{Me\kern-0.11em V}
\def\keV{ke\kern-0.11em V}

\begin{document}

\title{A deep wide survey of faint low surface brightness galaxies in the
direction of the Coma cluster of galaxies
\thanks{Based on observations obtained at the Canada-France-Hawaii Telescope 
(CFHT) which is operated by the National Research Council of Canada, the 
Institut National des Sciences de l'Univers of the Centre National de la 
Recherche Scientifique of France, and the University of Hawaii.}}

\offprints{C. Adami \email{christophe.adami@oamp.fr}}

\author{ C. Adami\inst{1} \and
R. Scheidegger\inst{2} \and
M. Ulmer\inst{2} \and
F. Durret\inst{3,4} \and
A. Mazure\inst{1} \and
M.J. West\inst{5} \and
C.J. Conselice\inst{6} \and
M.~Gregg\inst{7} \and
S.~Kasun\inst{2} \and
R.~Pell\'o\inst{8} \and
J.P. Picat\inst{8}
 }

\institute{
LAM, Traverse du Siphon, 13012 Marseille, France
\and
Department of Physics and Astronomy, Northwestern University,
2131 Sheridan Road, Evanston IL 60208-2900, USA
\and
Institut d'Astrophysique de Paris, CNRS, Universit\'e Pierre et Marie Curie,
98bis Bd Arago, 75014 Paris, France
\and
Observatoire de Paris, LERMA, 61 Av. de l'Observatoire, 75014 Paris, France
\and
Department of Physics and Astronomy,  University of Hawaii,
200 West Kawili Street, LS2, Hilo HI 96720-4091, USA
\and
Department of Astronomy, Caltech, MS 105-24, Pasadena CA 91125, USA
\and
Department of Physics, University of California at Davis, 1 Shields Avenue,
Davis, CA 95616
\and
Observatoire Midi-Pyr\'en\'ees, 14 Av. Edouard Belin, 31400 Toulouse, France
}

\date{Accepted . Received ; Draft printed: \today}

\authorrunning{Adami et al.}

\titlerunning{Low surface brightness galaxies in the Coma cluster}

\abstract
% context heading (optional)
{}
% aims heading (mandatory)
{We report on a search for faint (R total magnitude fainter than 21)
and low surface brightness galaxies (R central surface brightness
fainter than $\sim$24) (fLSBs) in a 0.72$\times$0.82~deg$^2$ area
centered on the Coma cluster. }
% methods heading (mandatory)
{We analyzed deep B and R band CCD imaging obtained using the CFH12K
camera at CFHT and found 735 fLSBs.  The total B magnitudes, at the
Coma cluster redshift, range from $-13$ to $-9$ with B central surface
brightness as faint as 27 mag arcsec$^{-2}$. }
% results heading (mandatory)
{Using empty field comparisons, we show that most of these fLSBs are
probably inside the Coma cluster. We present the results of comparing
the projected fLSB distributions with the distributions of normal
galaxies and with known X-ray over densities. We also investigate their
projected distribution relative to their location in the color
magnitude relation.  Colors of fLSBs vary between B$-$R$\sim$0.8 and
$\sim$1.4 for 2/3 of the sample and this part is consistent with the
known CMR red-sequence for bright (R$\leq$18) ellipticals in Coma.}
% conclusions
{These fLSBs are likely to have followed the same evolution as giant
ellipticals, which is consistent with a simple feedback/collapse
formation and a passive evolution. These fLSBs are mainly clustered
around NGC~4889. We found two other distinct fLSB populations. These
populations have respectively redder and bluer colors compared to the
giant elliptical red-sequence and possibly formed from stripped faint
ellipticals and material stripped from spiral in-falling galaxies.  }
%\keywords{galaxies: clusters: individual (Coma)}


\maketitle

\section{Introduction}\label{sec:intro}

In the last three decades, surveys of the local universe have revealed
the presence of galaxies only a few percent brighter than the sky
background, known as Low Surface Brightness (LSB) galaxies. LSBs have
remained mostly undetected because galaxy detection is contaminated by
the brightness of the night sky.  Little is known about LSBs: their
origin, physical properties (e.g.  luminosity, colors, radius) and
number density remain enigmatic. Because of the fundamental difficulty
in detecting LSBs, it is also possible that some types of LSBs are
still unknown. A number of studies have been carried out to identify
LSBs and study their origin but no clear scheme of formation and
evolution has been favored (e.g. Binggeli et al. 1985; Schombert et
al. 1992; Bothun et al. 1993; Bernstein et al. 1995; Impey et
al. 1996; Sprayberry et al.  1996; Ulmer et al. 1996, hereafter U96;
Impey \& Bothun 1997; O'Neil et al.  1997; Kuzio de Naray et al. 2004;
Sabatini et al. 2005). The present paper is devoted to the search for
faint Low Surface Brightness galaxies (fLSBs) in a cluster environment
and to a discussion of their origin and properties.

LSB galaxies are commonly defined by a central surface brightness
fainter than 22 or 23 mag per square arcsec in the B band (cf. Bothun
et al. 1991). Throughout this paper, we are interested in fainter
objects with the following definition: galaxies fainter than R=21
(absolute R magnitude fainter than $\sim -14$ at the distance of the
Coma cluster), with radius larger than $\sim$0.6 arcsec (slightly less than
3~kpc) and with R central surface brightness fainter than $\sim 24$ mag
arcsec$^{-2}$. The radius corresponds here to the standard deviation
of the Gaussian fit to the surface brightness profile of the fLSBs (see 
Sections 2.3 and 2.4).

Further studies of fLSBs and their properties are needed for several
reasons:

First, according to Cold Dark Matter (CDM) models of hierarchical structure
formation (White \& Rees 1978, White \& Frenk 1991), there should be abundant
low-mass dark matter halos present in the Universe. These halos could develop
low luminosity stellar systems and be detected as low luminosity galaxies.
However, CDM theory overestimates the number of such detected dark halos:
observations have reported far fewer low luminosity galaxies than predicted by
simulations (e.g. Davies et al. 2004, but also see Kravtsov et al. 2004 for
possible alternative solutions). A simple explanation for this discrepancy is
that these numerous low luminosity galaxies exist but are too faint to be
detected. Because fLSBs are strongly dominated by Dark Matter (e.g. McGaugh et
al. 2001, de Blok et al. 2001) and are by definition the most difficult low
luminosity galaxies to detect, they are the perfect candidate to fill the
apparent lack of low luminosity structures.

Second, another difference between CDM theory and observations is the
so-called ``dwarf to giant ratio" in different environments. According
to CDM models, low luminosity galaxies should be present and similar
in all environments. The rich galaxy clusters such as Coma, Fornax,
and Virgo have a substantial low luminosity population illustrated by
a high dwarf to giant ratio (e.g. Secker et al. 1997, Roberts et
al. 2004, Sabatini et al. 2005), but a possible lack of faint/dwarf galaxies
has been reported in lower density environments {\bf such as the Local
Group (e.g. Mateo 1998)}. To reconcile observations with theory, low
luminosity galaxies must be selectively destroyed or transformed in
low density environments and/or maintained or created in clusters of
galaxies. fLSBs being among the most sensitive galaxies to
environment-dependent processes, their studies in different
environments is crucial.

In this paper, we report on an extensive new study of  fLSBs in the Coma
cluster.  Rich environments can be harsh to dwarf galaxies and LSBs (e.g.
L\'opez-Cruz et al. 1997, Gregg \& West 1998).
Cluster galaxies can be affected by various processes that are not so
often at play for field galaxies: direct collisions, tidal interactions, high
speed encounters, ram pressure stripping by the intracluster medium (ICM), pressure
confinement and combinations of the above. Pressure confinement (e.g. 
Babul \& Rees 1992), however, does not work for galaxies moving through the
ICM at the typical velocity of rich clusters ($\geq 800$~km/s), where the
effects of ram pressure become important. 

The Coma cluster is one of the
densest nearby rich clusters and is therefore excellent for studying the effects
of environment on the formation and evolution of galaxies. It also has
the advantage of being located near the North Galactic pole, which makes the
effects of galactic absorption negligible. 

Coma has been extensively studied in the literature (see Biviano 1998 for a
review of works before 1995) and is a complicated cluster, with
evidence for several mergers (see recent reviews in "Merging Processes in
Galaxy Clusters" 2002, Feretti et al. Eds. Kluwer). It also contains two D (or
one cD and one D galaxies, see Schombert et al. 1992, L\'opez-Cruz et al. 1997),
X-ray emission with strong substructures (Neumann et al. 2003), an extended
radio halo (Giovannini et al. 1993) and a radio relic (e.g. Feretti \& Neumann
2006). Despite the wealth of observations on the Coma cluster, most works at
optical wavelengths were limited to relatively bright magnitudes (a few
examples are Andreon \& Cuillandre 2002, Beijersbegren et al. 2002,
Iglesias-P\'aramo et al. 2003, and Lobo et al. 1997) or to relatively small areas
with limited spectral coverage (e.g. Trentham 1998 with only two bands and 0.19
deg$^2$ or Bernstein et al. 1995 with a single deep band and 0.0145 deg$^2$). Our
data (Adami et al. 2006, hereafter A06) fill these gaps since they are at the
same time wide (0.72x0.82 deg$^2$ or 1.8 Mpc$^2$), deep (R$\sim 24$) and with a large
wavelength coverage (B, V, R, and I bands with equivalent depths).

The paper is organized as follows: we
present the observations and the fLSB detection algorithm
in Section 2. In Section 3 we compare our galaxy sample with
other surveys. In Section 4, we discuss the fLSB colors. In Section 5
we discuss the spatial distribution of the fLSBs and its relation with
that of the giant galaxies. In Section 6 we discuss our results in
terms of some possible mechanisms that occur specifically in the
cluster and give our conclusions in Section 7.

We assume a distance to Coma of 95 Mpc, H$_0$ = 75 km s$^{-1}$
Mpc$^{-1}$, $\Omega _\Lambda =0.7$, $\Omega _m =0.3$, distance modulus
= 34.89, and therefore the scale is 0.44 kpc arcsec$^{-1}$. All
magnitudes are given in the Vega system.

\section{Observations and data analysis}

\subsection{Observations}

The observations are described in A06 and we only reproduce the
salient points in this section. The Coma field was observed in April
1999 and April 2000 with the Canada-France-Hawaii 3.6m telescope using
the CFH12K camera. This camera is a mosaic of twelve CCDs. Two sets of
images were taken in order to cover both the north and south regions
of the cluster. The entire observed field covers an area of
0.72$\times$0.92 deg$^2$ centered on the two giant elliptical galaxies
NGC~4874 and NGC~4889 at the core of the Coma cluster. Images were
acquired using B, V, R and I Johnson-like filters. The seeing ranged
from 0.9~arcsec in R to 1.07~arcsec in B. One pixel corresponds to
0.206~arcsec.

Because the goal of this work was to detect fLSBs in the deepest band (R)
and then to investigate the distribution of fLSBs in a color magnitude
relation, we only used the R and B bands. We will use other bands
(including U band data that we plan to acquire) to study
spectrophotometric properties of fLSBs in a future work.

The data reduction specific to this project was a three step process: 
step 1 was to search for all the objects
in the R image using SExtractor (Bertin $\&$ Arnouts 1996); step 2
was to select all the faint low surface brightness objects from the
SExtractor catalog; and, step 3 was to derive the magnitudes for the
selected fLSBs in both R and B bands and to derive colors from
photometry in the same aperture.

\begin{figure}[bht]
\centering
\mbox{\psfig{figure=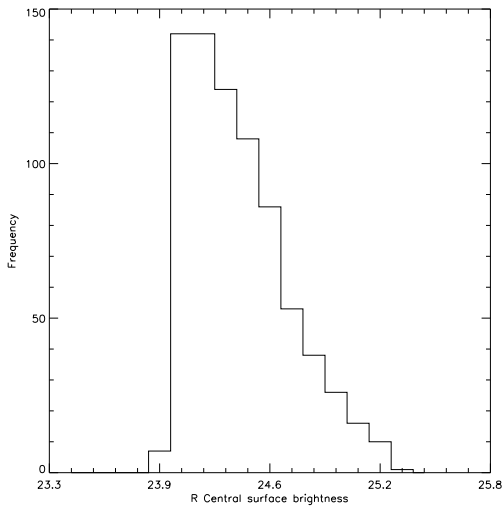,height=7.cm,angle=0}}
\caption[]{Histogram of the  central surface brightness values  of the
fLSBs. }
\label{fig:fig2}
\end{figure}

\begin{figure}[bht]
\centering
\mbox{\psfig{figure=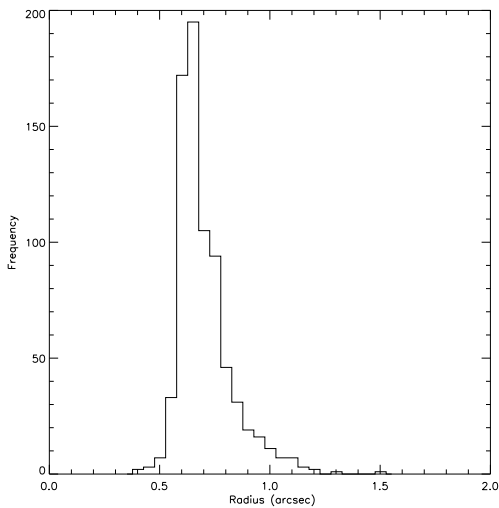,height=7.cm,angle=0}}
\caption[]{Histogram of the radius of all the selected fLSBs.}
\label{fig:fig3}
\end{figure}

\subsection{Step 1: Detection of sources}

We used SExtractor to detect all objects on the R
images, classify them as stars or galaxies and calculate among other things their
total magnitude, core magnitude and coordinates. Object detection with
SExtractor was not optimized to detect low surface
brightness objects. The resulting
SExtractor catalog contains over 60000 detections, including stars,
globular clusters, galaxies, etc.

\subsection{Step 2: Identification of low surface brightness galaxies}

The second step of the analysis was to identify fLSBs among all the objects in the
SExtractor catalog. The galaxies we are considering to be fLSBs
are not classical dwarf galaxies but the much fainter objects defined in the
introduction.  We distinguished fLSBs from other object types by applying a series of
selection criteria.

- First, we selected objects fainter than R=21.

This selection criterion for fLSBs is based on the properties of tidal dwarf galaxies
because fLSBs could possibly be tidal dwarfs. Tidal dwarf galaxies have masses
between 10$^7$ and 10$^8$M$_\odot$ (Bournaud et al. 2003). Assuming a M/L ratio in R
of about 5 (Mateo, 1998) for the most massive dwarf galaxies, their apparent
magnitude     should be fainter than R$\sim$21 at the Coma cluster redshift. This
defines the brightest magnitude cut for fLSB selection. It is also in  good agreement
with the LSB selection criteria used in U96. 

- Second, we identified fLSBs according to the shape of their surface brightness
profile. 

We differentiated fLSBs from other faint objects using Gaussian radial
surface brightness profiles. Although fLSBs typically have
exponential surface brightness profiles,  
U96 found that fLSB selection based on exponential profiles
generates a large number of false candidates in the rich environment
of Coma, due to the proximity of neighboring objects. In our data, exponential
fits are good when confined to the inner regions of the galaxies, but
because Gaussian fits are less sensitive to crowding they give better
fits up to the outer regions of the galaxies. Instead of using
exponential profiles, U96 proposed to select fLSBs by $\chi
^2$-fitting of Gaussian curves to the radial surface brightness
profiles of fLSBs. As shown in Section 2.4, this does not mean
that an exponential is not the proper form of fLSB profile. Rather,
the Gaussian profile is the result of the intrinsic (exponential)
shape convolved with instrumental effects (PSF, seeing). Then, following
U96, we used Gaussian fits to carry out the initial fLSB selection.

We fit a Gaussian form plus a constant background to the linear-scale surface
brightness profiles on the R image.  Initially, we let the radial profiles extend
to a radius $\theta _{max}$ = 2.5~arcsec from the center of each object, which, as
determined by visual inspection, encompasses the entire range of fLSB sizes.

- Third, we selected $initial$ fLSB candidates with radius greater than 0.6~arcsec
  and R central surface brightness fainter than $\mu _{\rm R}$ = 24 mag
  arcsec$^{-2}$. The size threshold was chosen above the seeing radius 
  in order to limit contamination by globular clusters which at the
  distance of Coma, appear as point sources. 

- Fourth, we optimized the fit parameters for all initial candidates and selected
all candidates with acceptable fits.

We optimized the final value of $\theta_{max}$ for all the selected     candidates
($\sim$1100) to ensure that none of their surface brightness profiles were
contaminated by surrounding objects. The optimized $\theta_{max}$ for each candidate
was determined by visual inspection. The fitting procedure was repeated. After
inspecting all candidates visually we selected as final fLSBs all the candidates that
yielded an acceptable (the probability of finding a larger $\chi ^2$ value is smaller
than 10\%) gaussian fit  to a distance of $\theta _{max}$. The resulting sample
contains 735 fLSBs.

After selecting the fLSB sample, we checked that their inclinations
did not introduce any bias in the selection procedure. If fLSBs are
disk-like, we might expect highly inclined fLSBs to have their surface
brightness artificially increased, making their detection
easier. However, we checked that high central surface brightness
objects are not systematically highly elliptical. The fLSBs with the
faintest central surface brightnesses have an ellipticity 8\% smaller
than the fLSBs with the brightest central surface brightnesses. 
Since this is smaller than the uncertainty in the magnitudes,
this effect is negligible.

In what follows, we will refer to $\sigma$, the standard deviation of
the Gaussian fit, as the fLSB radius except when explicitly noted.
Note that the final values of $\sigma$ and R central surface
brightness changed from our initial cutoffs because of the
optimization on $\theta _{max}$ performed after the initial candidate
selection.  For example, when $\theta _{max}$ was diminished in order
to avoid pollution by neighboring objects, the fit value of $\sigma$
changed because the brightness profile was also modified when the
neighbor was removed. Nevertheless, Figs.~\ref{fig:fig2} and
\ref{fig:fig3} show that most of the fLSBs still fall within the
original selection criteria.

\subsection{Point Spread Function effects on fLSB profiles}

As mentioned in the previous section, fLSBs typically should have
exponential profiles. However, Gaussian profiles provide a better fit to the
data. This is due to seeing effects. To demonstrate this, we first
convolved an exponential profile with scale length of 1.4 arcsec (the mean
scale factor from U96) with the average Gaussian fits of the point spread
function (PSF) in our data. We found the net result to have 
a shape that is better fit by a Gaussian than an exponential (see also  
Fig.~\ref{fig:seeing2}).  

We further investigated the effects of seeing variations on an
exponential profile across the field of view.  To quantify the
instrumental effects on the PSF across the Coma field, we derived the
seeing in 100 sub-regions by fitting 2D elliptical Gaussian profiles
on $\sim$800 known stars between magnitudes I=18.5 (to avoid
saturation) and I=20.5 (to avoid confusion with compact galaxies, see
A06). We then computed the orientation of the major axis of the stars
and the FWHM along the major and minor axes. This was done on the same
R band images used for the fLSB detection.

\begin{figure} 
\centering
\mbox{\psfig{figure=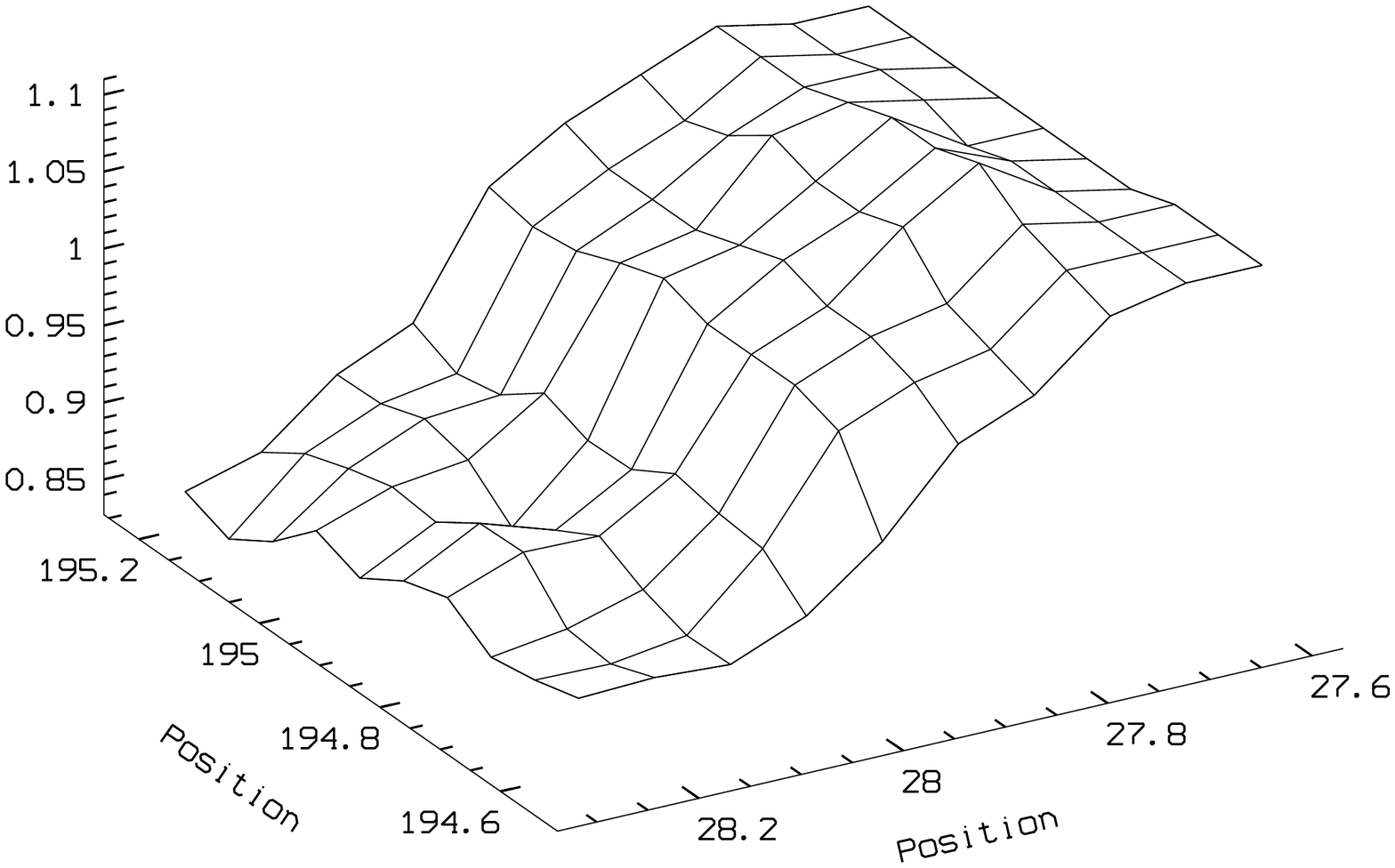,height=8.cm,angle=270}}
\mbox{\psfig{figure=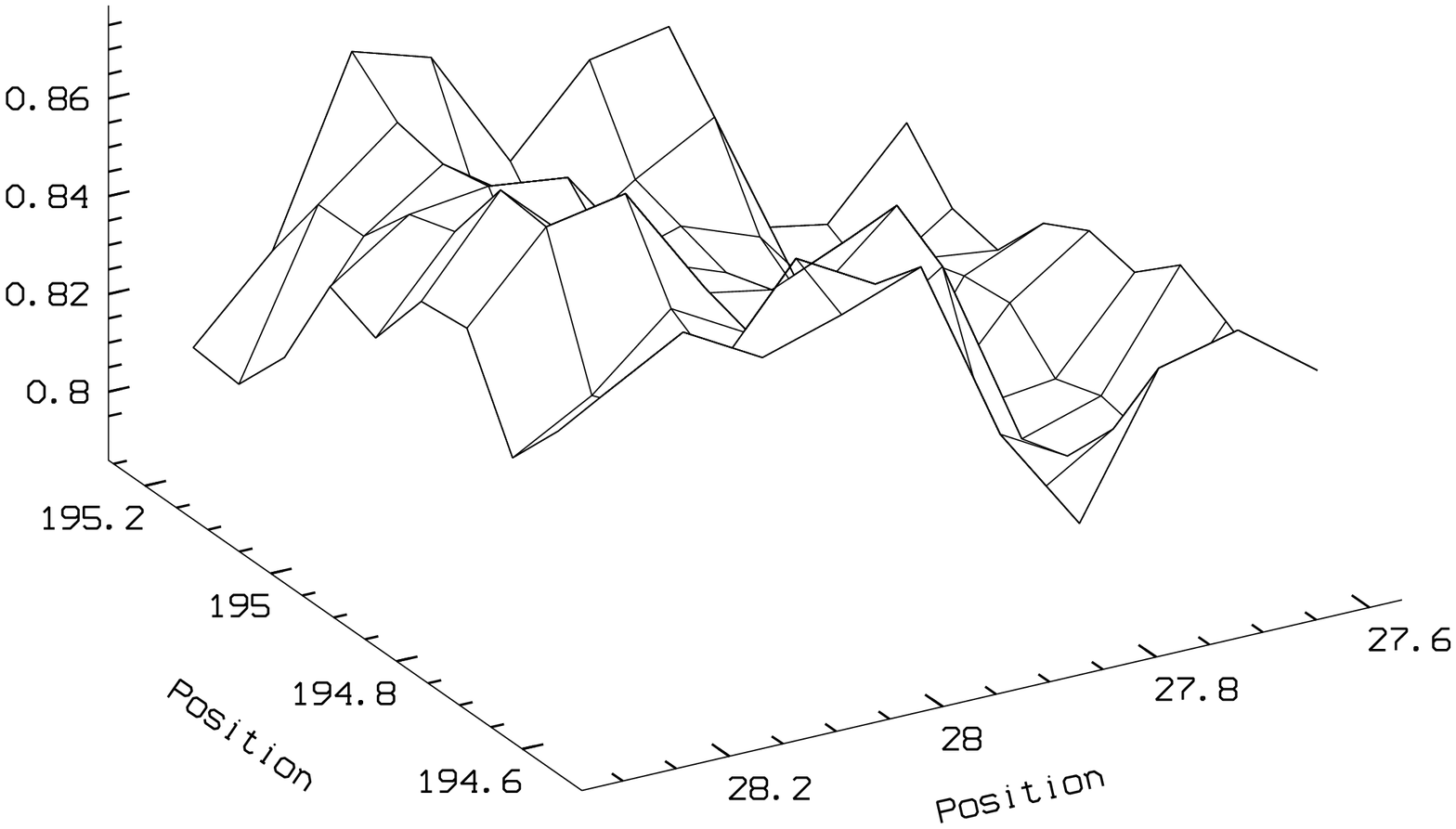,height=8.cm,angle=270}}
\mbox{\psfig{figure=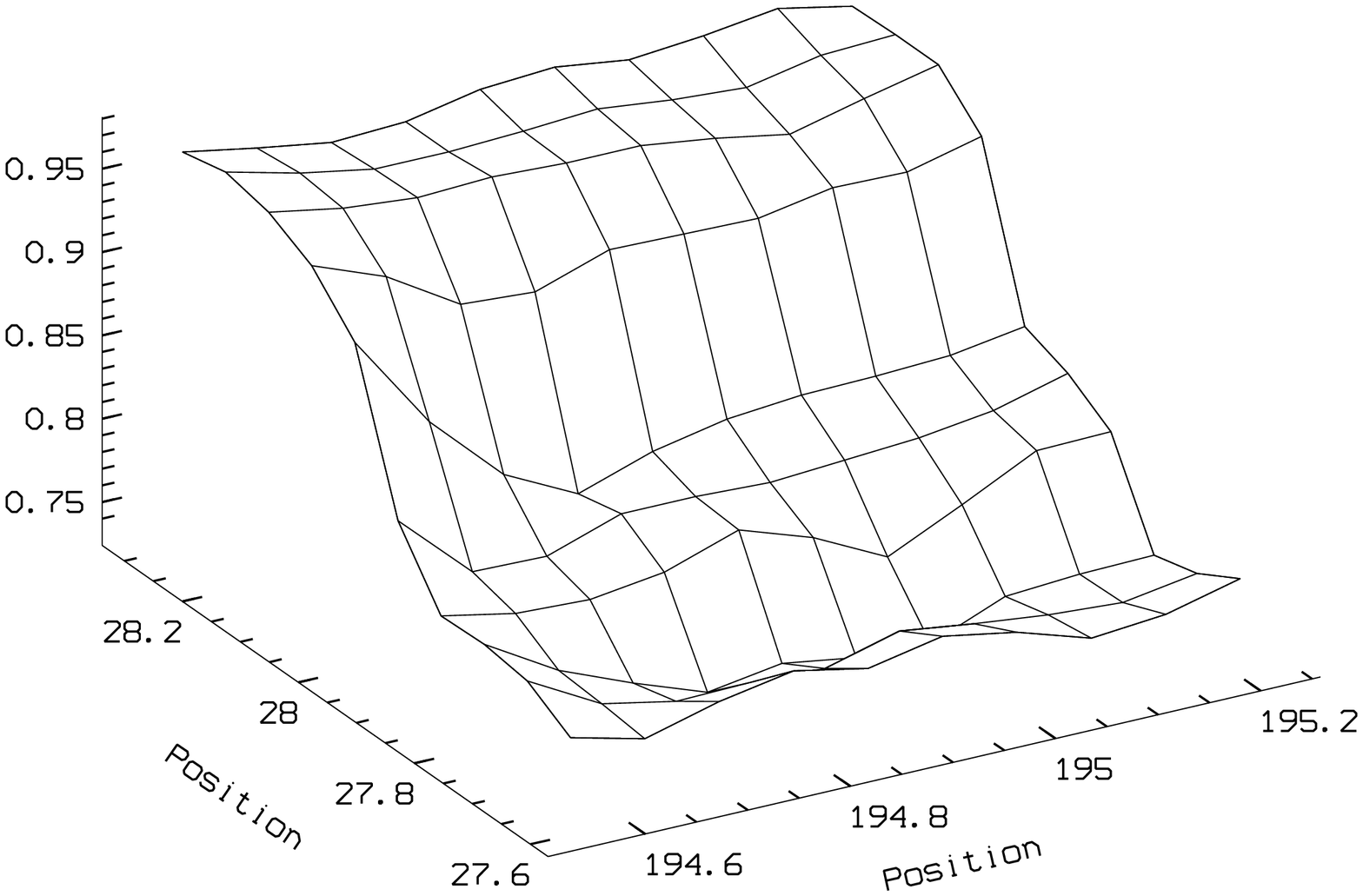,height=8.cm,angle=270}}
\caption[]{3-D representation of the PSF FWHM along
the major axis (top) and minor axis (middle), and
3-D representation of the ratio of the minor axis FWHM
to the major axis FWHM (bottom). $\alpha$ and $\delta$ are given in
decimal degrees.}
\label{fig:seeing1_1}
\end{figure}

\begin{figure} 
\centering
\mbox{\psfig{figure=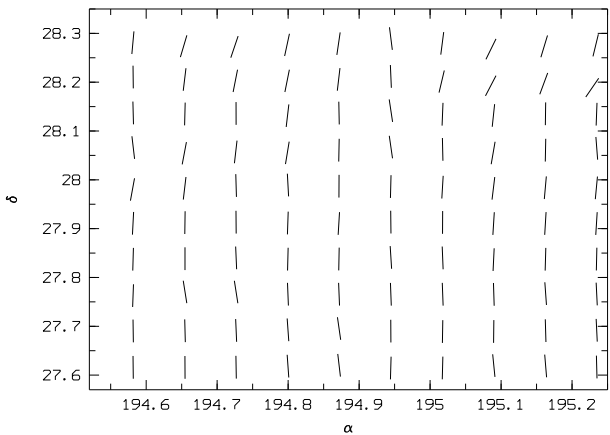,height=6.cm,angle=270}}
\caption[]{Map of the major axis orientation. $\alpha$ and $\delta$
are given in decimal degrees (note that $\alpha$ increases to the
right).}
\label{fig:seeing1_2}
\end{figure}

\begin{figure} 
\centering
%\mbox{\psfig{figure=convolveRfirstnord.ps,height=6.cm,angle=270}}
%\mbox{\psfig{figure=convolveRfirstsud.ps,height=6.cm,angle=270}}
\mbox{\psfig{figure=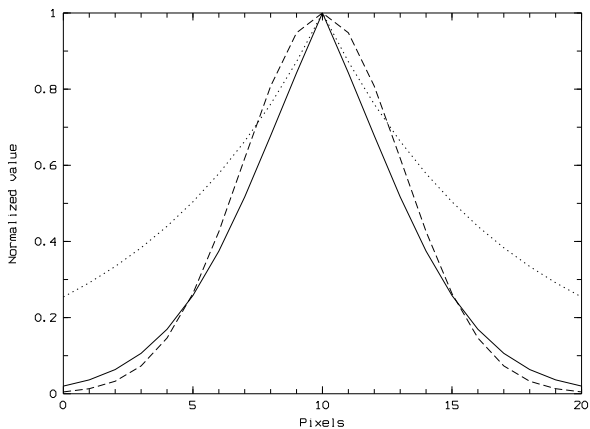,height=6.cm,angle=270}}
\caption[]{Plots for the whole Coma cluster region. Small-dotted line:
original exponential profile. Continuous line: original
exponential profile convolved by the PSF. Dashed line: mean
fLSB profile observed in our data. All curves have been normalized to
the same maximal value. Pixels are 0.205 arcsec.}
\label{fig:seeing2}
\end{figure}

We then computed and smoothed these maps to produce
Figs.~\ref{fig:seeing1_1} and ~\ref{fig:seeing1_2}. These maps show the 
average of these PSFs, and the ratio between the FWHM of the
minor and major axes. We also averaged the orientation of the major axis.
We carried out the averaging over each artificial pixel 
($\sim 0.07 ^\circ \times 0.08 ^\circ$ in size),
using an adaptive kernel technique (e.g. Adami et al.  1998). 

We clearly see on these maps an elongation of the point spread fucntion (PSF) along the
$\delta$-direction (except in the north-east area). It is larger in the
southern region (FWHM close to 1.05 arcsec) than in the north (FWHM
close to 0.9 arcsec) while the minor axis is relatively constant (FWHM
close to 0.83 arcsec) on the whole field. This results in a
major/minor axis ratio of more than 0.9 in the north and less than 0.8
in the south, perhaps partially due to bleeding in the CCD readout.

To examine the effect of such a PSF on the observed profiles, we
convolved an exponential profile using the mean scale factor from 
U96 (a scale of 1.4 arcsec) with the
average Gaussian fits of the PSF in our data.

The results shown in Fig.~\ref{fig:seeing2} imply that 
the convolved profile is well fit by a Gaussian and we also checked that the 
difference
in PSF between the north and south regions only has a minor effect. 
This justifies
the use of a Gaussian fit to find the fLSBs with no necessity to treat the
north regions differently from the south.

\subsection{Step 3: Colors and total magnitudes.}

In Step 3 we computed the total magnitudes and colors of the 735 fLSBs we detected in
Step 2. Total magnitudes were calculated from the radial surface brightness profiles
while colors were derived from common aperture photometry.

We calculated total magnitudes by integrating to infinity the Gaussian radial surface
brightness profiles generated in the previous step and converting the total counts to
magnitudes. Following the discussion in \S 2.4, we used the Gaussian profiles rather
than the exponential profiles because the Gaussian profiles describe the observed
data more accurately. The central surface brightness values were derived by
extrapolating the Gaussian fits. Inspection of the profiles in Fig.~\ref{fig:fig4}
suggests that the central surface brightness values are reliable.

We measured colors by determining B and R magnitudes for each fLSB within the same
aperture. The aperture size was adapted to the radius and crowding of each fLSB.  The
aperture radius was taken to be the final $\theta _{max}$ (see \S 2.2), which was
also the inner radius of the local sky background annulus.  The outer radius of the
local sky background annulus was set to 4.5~arcsec, except in case of overlapping
neighboring objects. When close neighbors were present, the outer radius of the local
sky was set to the maximum possible radius free of other objects detectable above the
background.

\subsection{Total magnitude error estimates}

\begin{table*}
\caption{Col.~1: fLSB identification numbers; Cols.~2-3: RA and Dec
(equinox 2000); Cols.~4-5: total magnitude in R and B based on
integrating the Gaussian fits to infinity for the R and B images;
Col.~6: B$-$R color derived using a fixed aperture photometry;
Cols.~7-8: central surface brightness for R and B derived by
interpolating the best Gaussian fits to r=0; Cols.~9-10: radius in arc seconds
for R and B. This table only shows the first 5 fLSBs, the full
table can be downloaded from
$ http://www.astro.northwestern.edu/$$\sim$$ulmer/private/coma/lsb-table.txt $
or $ http://cencosw.oamp.fr/ $}
\begin{tabular}{llllllllll}
\hline
ID & Ra &  Dec &  M(R) &  M(B) &  B$-$R & $\mu_0$(R) & $\mu_0$(B) & $\sigma$(R) &
$\sigma$(B) \\
\hline
1  &  12:58:30.59 &  28:22:56.70 &   22.90 &  23.75  &  0.89 &  24.19 &
   25.30 &  0.72  &  0.88  \\
2  & 12:58:10.47 &   28:22:51.30 &  22.70 &  24.05  &  1.37 &  24.08 &
   24.95  & 0.77  &  0.46   \\
3  & 12:58:16.78  &  28:19:46.80 &  23.09 &  24.22  &  1.15 &  24.27 &
   25.49 &  0.59  &  0.74   \\
4  & 12:58:13.19  &  28:19: 9.70 &  22.89 &  23.74  &  0.84 &  24.38 &
   25.32 &  0.75 &   0.80   \\
5  & 12:58:28.60  &  28:19: 7.50 &  22.81 &  24.00  &  1.29 &  24.18 &
   25.02 &  0.78  &  0.34   \\
\hline
\end{tabular}
\label{tab:tab1}
\end{table*}

We investigated magnitude uncertainties by comparing the magnitudes of fLSBs detected
twice. There is a $\sim$7~arc-min overlap band between the two sets of images acquired
to cover the entire cluster field. For fLSBs detected in both image sets, the total
magnitude differences between both detections are less than 0.3 mag, in good
agreement with the uncertainties found by A06.

Typical uncertainties for the colors are 0.35 mag at R$\sim$24 and 0.15 at R$\sim$20
(estimated from the quadratic sum of individual magnitude uncertainties given in
A06).

\subsection{Comparison with SExtractor total magnitudes}

Fig.~\ref{fig:sexnew} shows that, for the very peculiar object class
investigated here, SExtractor total magnitude estimates can differ
from the analysis used to derive magnitudes of fLSBs, especially for
the brightest fLSBs (contrary to normal galaxies in Coma:
e.g. A06). We also found the same behavior in the B band data. The
systematic offset is possibly due to a source confusion effect as the
brighter fLSBs are also larger.  Given the uncertainties in the
derived magnitudes, however, the difference between the SExtractor
magnitudes and ours does not strongly influence our conclusions.

\begin{figure*} 
\centering
%\mbox{\psfig{figure=r2.ps,width=17cm,angle=0}}
\caption[]{Six sample plots  showing the R radial intensity profiles of
the fLSBs we detected, along with the best Gaussian (dotted curve) and
exponential (dashed  curve) fits. The bottom  X-axes are in pc. The top X-axes
are in arc seconds. The Y-axes are in R mag per square arc seconds. The
exponential fits show
some curvature because these fits (as well as the Gaussian model) include a
term for the background.}
\label{fig:fig4}
\end{figure*}

\subsection{The Coma field fLSB sample: summary}

We found 735 faint, low surface brightness galaxies in the direction
of the Coma cluster with central surface brightnesses ranging from
$\sim$24 to 25.5 R mag arcsec$^{-2}$ (except for 10 fLSBs which are
brighter) and from 24 to 27 B mag arcsec$^{-2}$. The total magnitudes
range from $\sim$21 $\leq$ R $\leq$ 24.5 and $\sim$22 $\leq$ B $\leq$
26. The colors are distributed for most fLSBs between 0.2 $\leq$
B$-$R$\leq$2.6 and peak at 1.2. fLSB radii range in $\sim$0.4 $\leq$
$\sigma$ $\leq$ 1.5~arcsec corresponding to $\sim$0.17 $\leq$ $\sigma$
$\leq$ 0.66 kpc but the majority have radii between $\sim$0.6$\leq$
$\sigma$ $\leq$ 0.8~arcsec or $\sim$0.26 $\leq$ $\sigma$ $\leq$ 0.35
kpc.  After quadratically subtracting the seeing radius value to these
numbers, most of the fLSBs fall in the range [0.20,0.31]~kpc.
In contrast, globular clusters have a half light radius (with a partially
different radius definition) of $\sim$0.003
kpc (e.g. van den Bergh et al. 1991 or Jord\'an et al.  2005).

Although the fLSBs we detected along the Coma line of sight
are very small and faint, they are certainly
not spurious since all were found on the R images but also on
the B images. Because the SExtractor detection
thresholds in R were quite
stringent (detection threshold of 2$\sigma$ and minimum number
of pixels of 9 above threshold (see A06)), a number
of fLSBs were therefore missed in R, but the detected ones
were quite obvious and then also detectable in other bands.

\begin{figure} 
\centering
\mbox{\psfig{figure=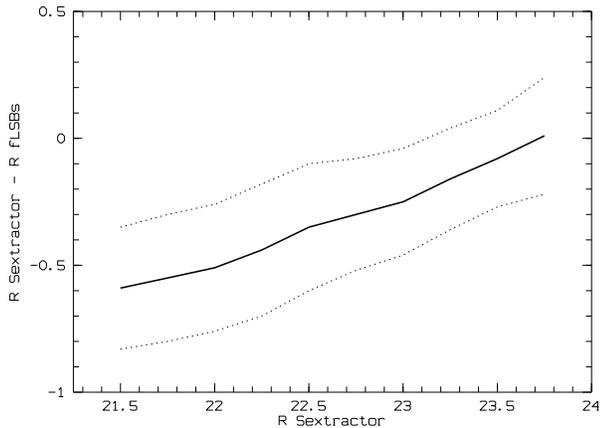,height=6cm,angle=270}}
\caption[]{Thick line: mean difference between SExtractor total R
magnitudes and present integrated magnitudes versus SExtractor total
magnitudes for the R band. Dotted lines: 1$\sigma$ error envelope.}
\label{fig:sexnew}
\end{figure}
\begin{figure} 
\centering
\mbox{\psfig{figure=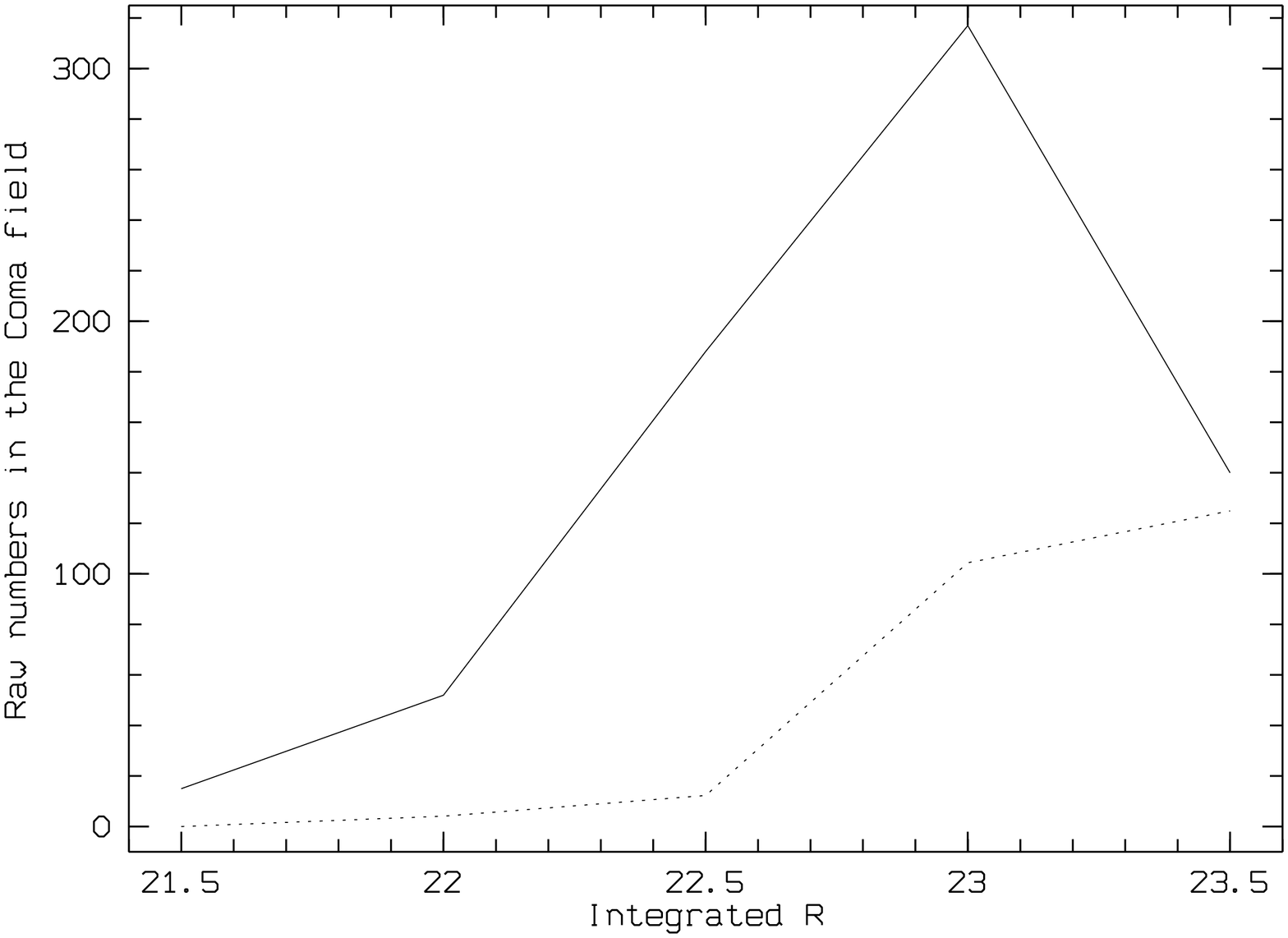,height=6.cm,angle=270}}
\mbox{\psfig{figure=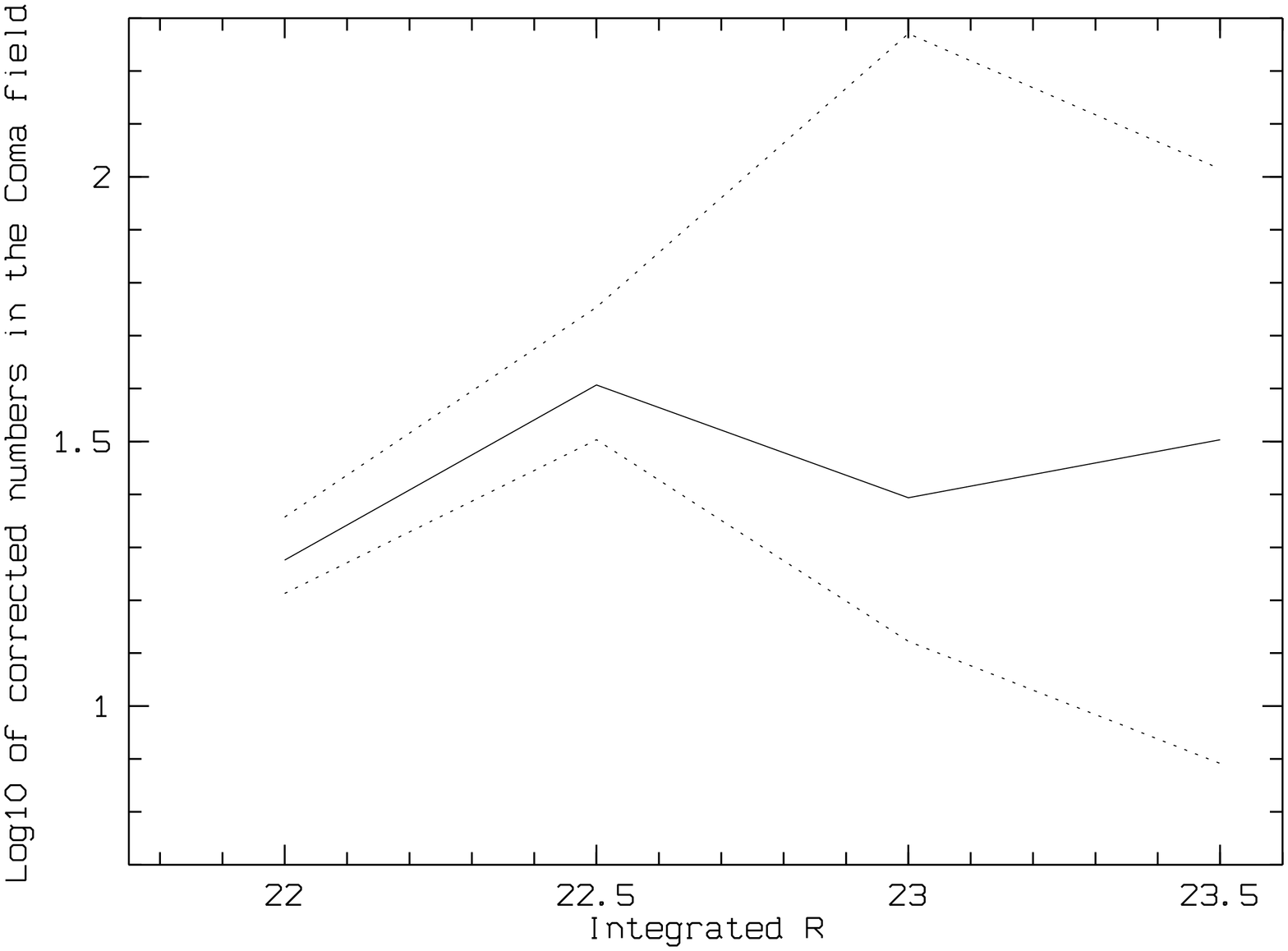,height=6.cm,angle=270}}
\caption[]{Upper figure: without detection efficiency corrections: the
solid line is the raw number of fLSBs detected along the Coma cluster
line of sight; the dotted line is the raw number of fLSBs detected
along the empty field line of sight scaled to the Coma cluster field
size. Lower figure: corrected for detection efficiency: Log10 of the
ratio between fLSBs detected in the Coma cluster field and fLSBs
detected in the empty field scaled to the Coma cluster field size. The
continuous line is the mean value, the dotted lines are the ratios
that delimit $\pm$10$\%$ uncertainty in the detection efficiency
estimates.}
\label{fig:figfield}
\end{figure}

Our fLSBs are similar in terms of size and total brightness
to those found in other clusters by the most recent fLSB searches.
They most closely resemble those found in Virgo by Sabatini et
al. (2005) with central surface brightnesses of B$\sim$26
arcsec$^{-2}$ and absolute B mag of about $-10$, and those found in
Ursa Major by Roberts et al. (2004) with a central surface brightness
average of 24.5 mag arcsec$^{-2}$ and scale lengths between 0.23 and
0.35~kpc. The fLSB colors are also in good agreement with the expectations
from Conselice et al. (2003) who find, in the Perseus
cluster, B$-$R colors ranging from 0.7 to 1.9, with a mean of 1.15 for
galaxies as faint R$\sim$21.3 at the Coma cluster redshift.
A small portion of the catalog of our results is given in
Table 1 where the web page address to the full catalog is given.

\subsection{Coma cluster membership}

We used two methods to investigate the cluster membership of our fLSB sample:
a  statistical comparison with an empty field  and an absolute magnitude
versus central surface brightness  comparison.

In order to put on a firmer ground the Coma membership of our fLSBs,
we need to estimate the number of foreground and background galaxies
satisfying our fLSB selection criteria. We therefore applied our fLSB
selection procedure to a 30$\times$30~arc-min$^2$ empty field extracted from 
the F02 field in
the deep VVDS R imaging survey (McCracken et al. 2003), observed with
the same instrument (CFH12K), the same R filter, and free from nearby
rich structures (VVDS collaboration, private communication). The
seeing values of these images are also similar to the seeing for Coma:
0.8 arcsec in R and 0.9 arcsec in B.

We corrected for detection efficiencies using McCracken et al. (2003)
for the VVDS R data and A06 for the Coma R data. This is shown in Fig.~\ref{fig:figfield}. 
We found that the number of
field galaxies satisfying the fLSB selection criteria are less than
4\% of the total number of fLSBs detected along the Coma line of
sight.

\begin{figure} 
\centering \mbox{\psfig{figure=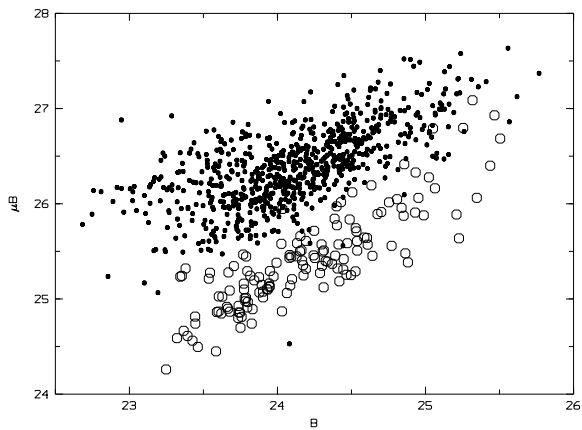,height=6.cm,angle=270}}
\caption[]{B magnitude versus
  B central surface brightness. Small dots: Coma line of sight fLSBs, open
  circles: empty field fLSBs.}
\label{fig:muBB}
\end{figure}

Another way to discriminate between cluster and line of sight fLSBs is the
absolute vs. surface brightness relationship.
In their work on the Perseus cluster, Conselice et al. (2002, see
their Fig. 6) have shown that the locations of background and cluster
galaxies in the absolute magnitude vs. surface brightness plane are
very different. We placed our
Coma line of sight and field fLSBs on a similar plot
(Fig.~\ref{fig:muBB}), which clearly shows that the location of most
of our fLSBs is inconsistent with the location of the empty field
fLSBs.  Only 5\% of fLSBs on the Coma line of sight overlap with the
empty field location, which is in good agreement with our previous estimate
and with the work by Conselice et al. (2002). 

These two methods show that the large majority of our fLSBs are likely 
Coma members and the foreground and background fLSBs are only a minor
contribution to the Coma sample.

\section{Comparison with other surveys}

To illustrate the sensitivity of our survey, we compare our sample
with previous fLSB catalogs in the core of the Coma cluster (U96) and
with the catalog of low surface brightness galaxies in Fornax (Bothun
et al.  1991).

\subsection{The Coma cluster core survey}

U96 conducted a survey of low surface brightness galaxies in the core
of the Coma cluster. The area they surveyed
($\sim$7.5$\times7.5$~arc-min$^2$) is centered just South of the
dominant galaxies and corresponds to portions of two CCDs in our
southern image. The catalog used by U96 to select fLSBs is currently
one of the deepest surveys of the Coma cluster with a completeness of
50\% down to R=25.5, but with a seeing close to 1.4 arcsec.  For
comparison, we applied our selection criteria to the U96 detections
(see Figs.~\ref{fig:fig2} and ~\ref{fig:fig3}) in terms of surface
brightness and minimal object size. We also limited the U96 sample to
the magnitude range 21$<$R$<$ 22.5 as, at the upper limit, our fLSB
detections are 50\% complete (see A06). We should then expect to
recover statistically in our sample about 50\% of the 7 U96 fLSBs
selected.  We did recover 3 fLSBs in our data, a value very close to
the 50\% expected level.

\subsection{Comparison with the Fornax cluster}

We also compared our fLSBs with the Fornax sample of Bothun et al.
(1991). 
The curved lines drawn on Fig.~\ref{fig:fig9} are derived by assuming
an exponential profile for the galaxies with different scale lengths
as indicated by the diagonal dashed lines ($\alpha$ being the
exponential scale factor in arcsec). The region right of each curved
solid line (toward lower B) is where we would expect to detect fLSBs
for each sample, given the angular diameter (smaller objects cannot be
distinguished from stars) and isophotal surface brightness limits (the
faintest level out to which an image is actually detected) noted next
to these curves. As expected, this figure shows that most of our fLSBs
fall to the right of the left-most curved line. If we correct for the
distance of Fornax, the brighter end of our fLSBs overlaps with the
faint end of the Fornax dwarfs. This demonstrates that our objects
are similar to the Fornax objects.

\begin{figure*} 
\centering
%\mbox{\psfig{figure=csbB_vs_B.ps,height=10.5cm,angle=90}}
\caption[]{Figure reproduced from U96 with the addition of our
data: central surface brightness as a function of total magnitude
in the B band. The pluses indicate Fornax galaxies from Bothun et
al. (1991), the asterisks are from U96, the diamonds are the U96
fLSBs put at the distance of Fornax (as denoted by the arrows),
the dots to the left are our fLSBs and the dots to the right are
our fLSBs put at the distance of Fornax. The solid curves show the
selection function that relates the limiting central surface
brightnesses and the limiting diameters of the objects to the
exponential scale factor given by B$_{tot} = -0.6689$ + 5log[($\mu
_{lim}$ - $\mu _0$)/$\theta _{lim}$] + $\mu _0$ where $\mu _{lim}$
is the limiting isophotal magnitude, $\mu _0$ is the observed
central surface brightness and $\theta _{lim}$ is the limiting
diameter in arc seconds. These values are only an estimate. The
values $\alpha$ are the exponential scale factor in arcsec.}
\label{fig:fig9}
\end{figure*}

\section{fLSB color analysis}

\subsection{Colors as a function of fLSB size}

\begin{figure} 
\centering
%mbox{\psfig{figure=MuR.ps,height=5.cm,angle=270}}
%\mbox{\psfig{figure=radiusR.ps,height=6cm,angle=270}}
\mbox{\psfig{figure=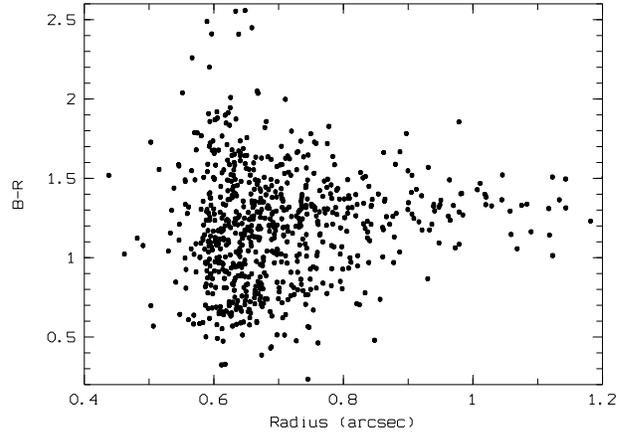,height=6cm,angle=270}}
\caption[]{B$-$R color vs fLSB radius.} \label{fig:fig10}
\end{figure}

In this section we examine how the fLSB colors vary with the radius
$\sigma$ (Fig.~\ref{fig:fig10}).  The smaller fLSBs have a large color
scatter while the larger fLSBs have a better-defined color sequence
centered around B$-$R$\sim$1.3.  We could expect such a wider spread
of colors for the smaller fLSBs since their lower binding energies
mean they are more likely to be affected by the cluster environment
(Grebel 2001) or more likely to lose metals produced by supernovae
(e.g. Kodama \& Arimoto 1997).  Alternatively, there could also be
more line of sight contamination toward the faint end of our sample,
since the number of line of sight objects increases with
magnitude. The color scatter of fLSBs is further investigated with the 
color-magnitude relation in the following sections.

\subsection{Color--magnitude relation}

One of the best known relations between the global properties of
galaxies and their stellar populations is the color-magnitude relation
(CMR).  Luminous early type galaxies in clusters are observed to be
redder than fainter ones. This progressive reddening of elliptical
galaxies with increasing luminosity is known as the CMR
red-sequence. The slope seen in the CMR red-sequence is driven
primarily by a luminosity-metallicity correlation (e.g. Kodama $\&$
Arimoto 1997 or Vazdekis et al.  2001): brighter galaxies have greater
binding energies and can therefore become more metal-rich and thus
redder than fainter ones.  The CMR red-sequence for the bright
galaxies in the Coma cluster has been well studied (e.g. Terlevich et
al.  2001, Odell et al. 2002, L\'opez-Cruz et al. 2004) and can be
compared to our fLSBs.

        L\'opez-Cruz et al. (2004) found the best fitting CMR red-sequence
        for Coma early type galaxies to be:

        B$-$R = $-$ 0.046 R + 2.22

\begin{figure} 
\centering \mbox{\psfig{figure=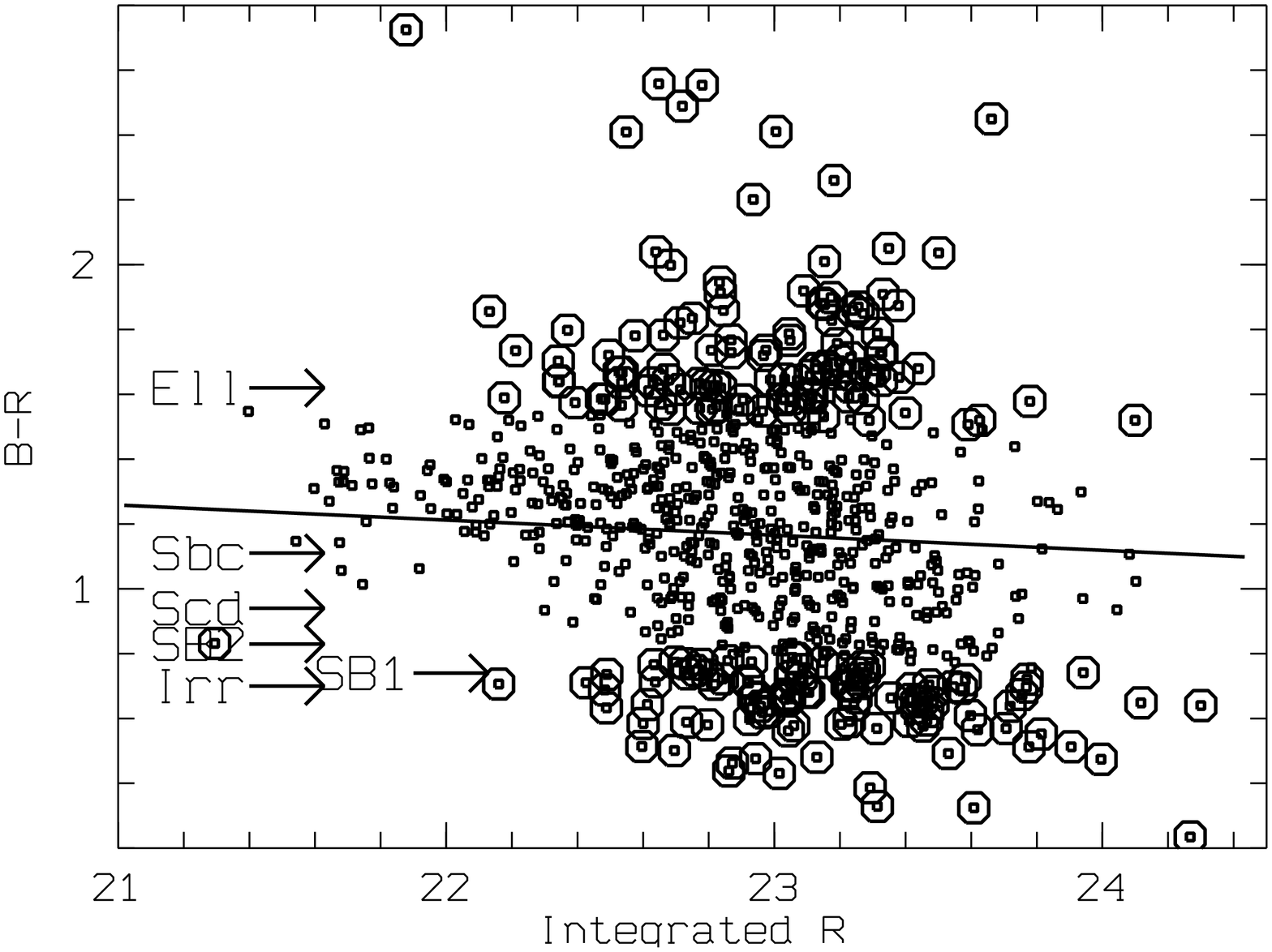,height=6.5cm,angle=270}}
%\mbox{\psfig{figure=CMRnew2.ps,height=6.5cm,angle=270}}
\mbox{\psfig{figure=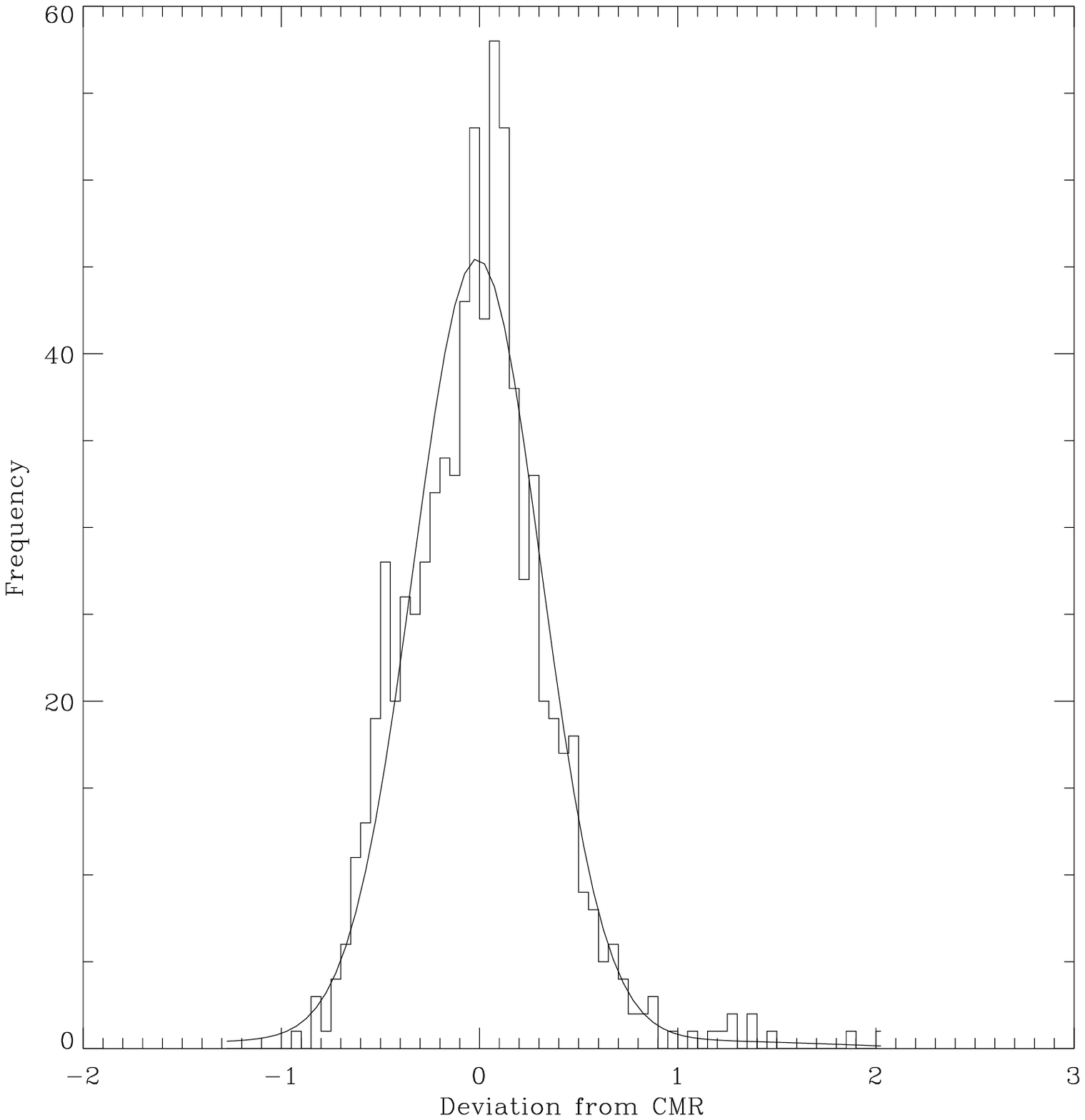,height=7.5cm,angle=0}}
\caption[]{Upper figure: color--magnitude relation for all the
fLSBs along with the CMR red-sequence given by L\'opez-Cruz et al.
(2004). Circled dots are fLSBs more than 1 standard deviation away
from the L\'opez-Cruz et al. (2004) relation. Theoretical colors for
several synthetic bright galaxy spectrophotometric types are also
quoted. Middle figure: fLSB B$-$R (filled circles) superimposed on
the whole object population B$-$R (small dots). Lower figure:
histogram of the deviations around the mean CMR red-sequence given
by L\'opez-Cruz et al. (2004), along with the best Gaussian fit on
this distribution.} \label{fig:fig11}
\end{figure}

This relation is consistent with most literature studies (e.g. Adami
et al. 2000, 2006) and, for comparison with our data, was corrected
for our specific B and R filters using the transformations given by
Fukugita et al. (1995). Results are shown in Fig.~\ref{fig:fig11} along
with the color magnitude diagram of our fLSBs. Surprisingly, the fLSB
colors are centered on the given CMR red-sequence for objects up to 10
magnitudes fainter than the brightest Coma elliptical galaxies!
L\'opez-Cruz et al. (2004) found a very narrow ($\sim$0.06 mag Gaussian
deviation) CMR red-sequence for the bright objects. Similarly,
we fit a Gaussian to the fLSB color distribution and corrected our
results for the intrinsic uncertainty on the magnitudes ($\sim$0.25 at
R=23, A06). We then find an intrinsic scatter of 0.27 mag around the
L\'opez-Cruz et al. (2004) CMR red-sequence, clearly larger than for the
bright objects but still significantly lower than for the whole object
distribution (Fig~\ref{fig:fig11}) which shows a 0.73 intrinsic
dispersion in the B$-$R [21,24.5] range. A Kolmogorov-Smirnov test
performed on the R/B$-$R relation shows that the two populations (the
whole sample and the fLSB sample) are different at the 99.99\% level
and confirms that our fLSBs are not a randomly selected sample from
the whole line of sight object population. We note that the star
contribution to the object counts in that magnitude range is
smaller than 10\%  and therefore negligible (A06, Bernstein et al. 1995).

\begin{figure} 
\centering \mbox{\psfig{figure=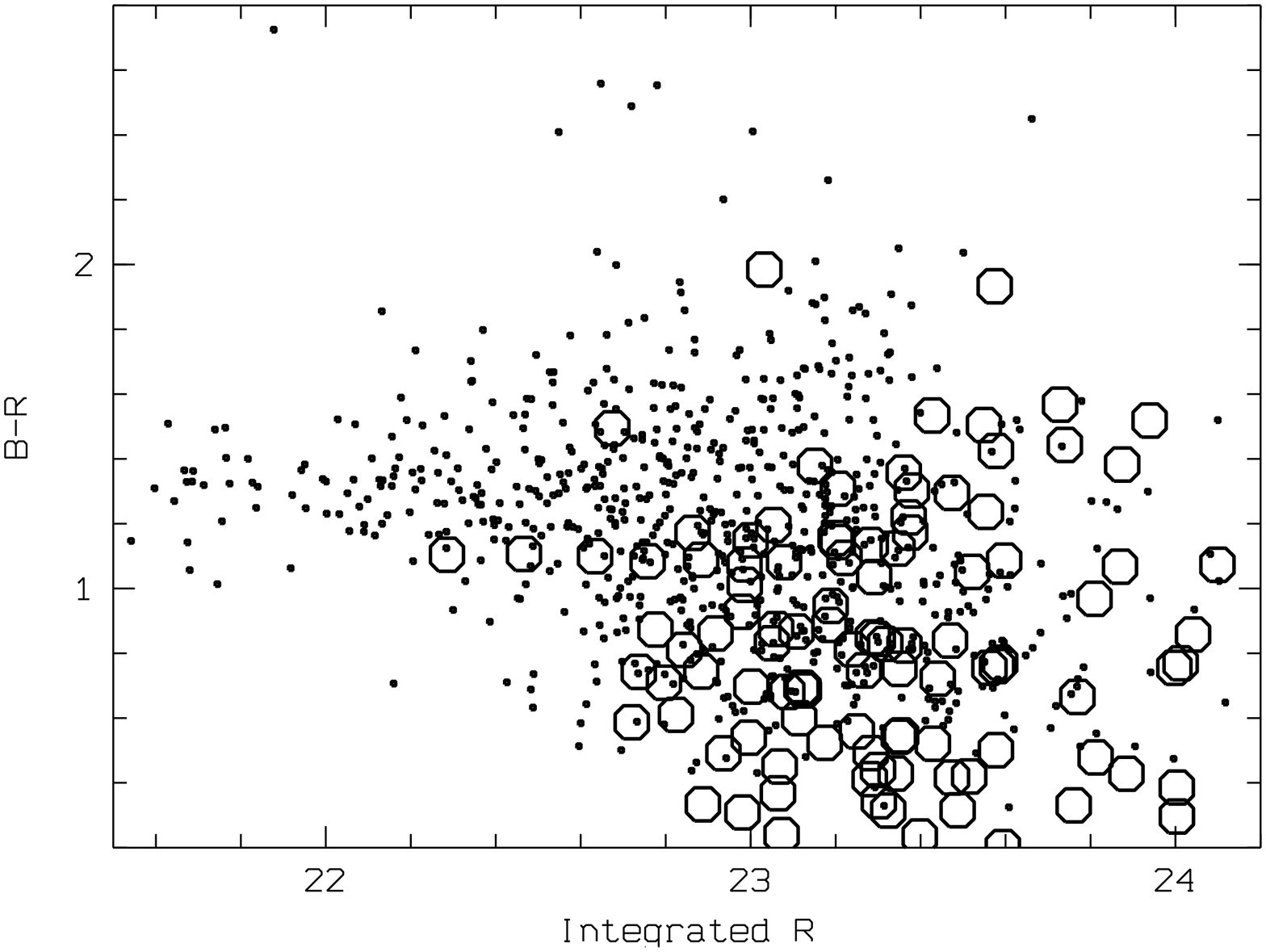,height=6.cm,angle=270}}
\mbox{\psfig{figure=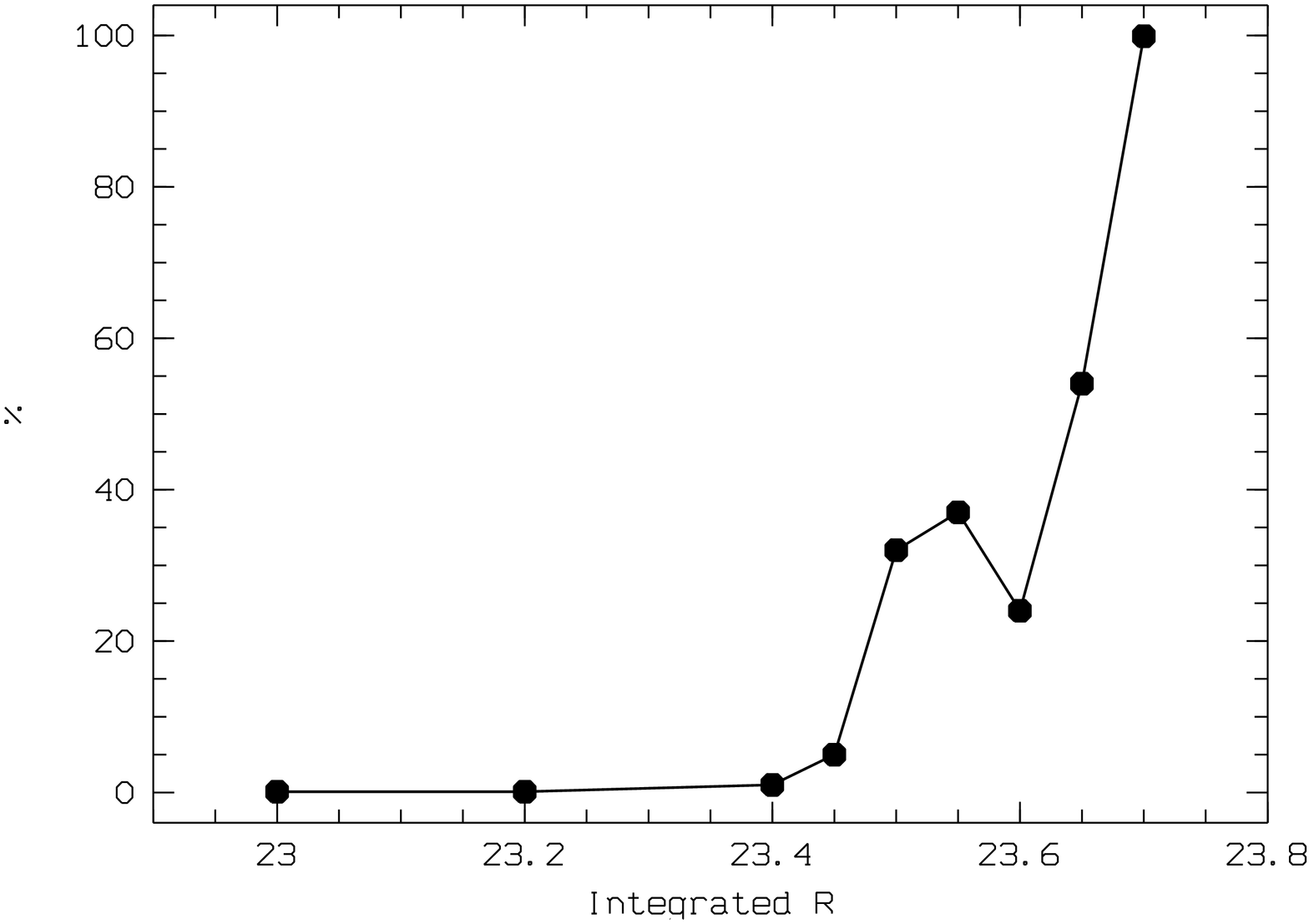,height=6.cm,angle=270}}
\caption[]{Upper graph: Small dots: fLSBs detected along the Coma
cluster line of sight. Large circles: fLSBs detected in the empty
field. Lower graph: Probability of the Coma and empty field fLSB
samples fainter than a given R integrated magnitude to be
similar.} \label{fig:figrepres}
\end{figure}

\begin{figure} 
\centering
\mbox{\psfig{figure=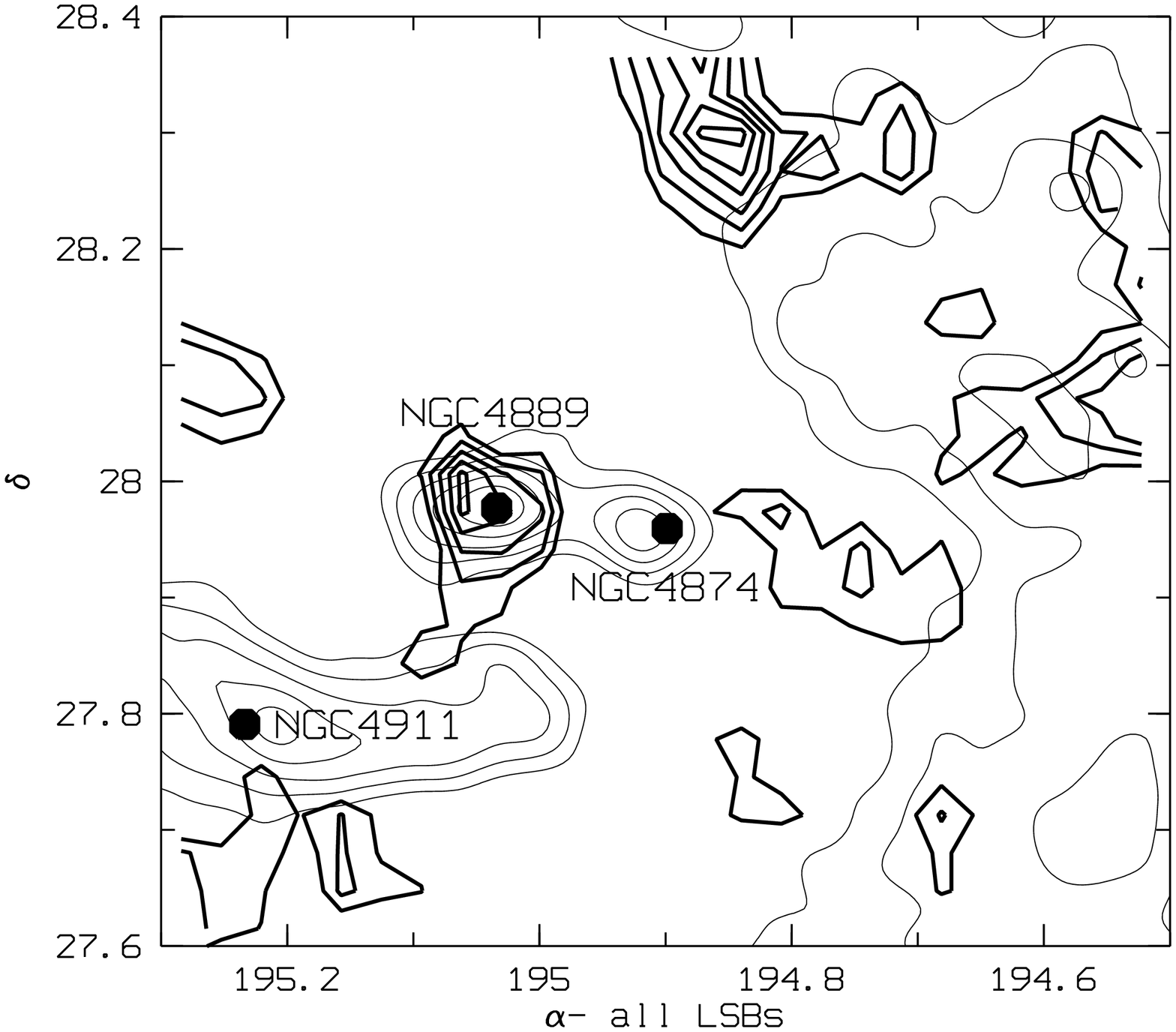,height=7.cm,angle=270}}
\mbox{\psfig{figure=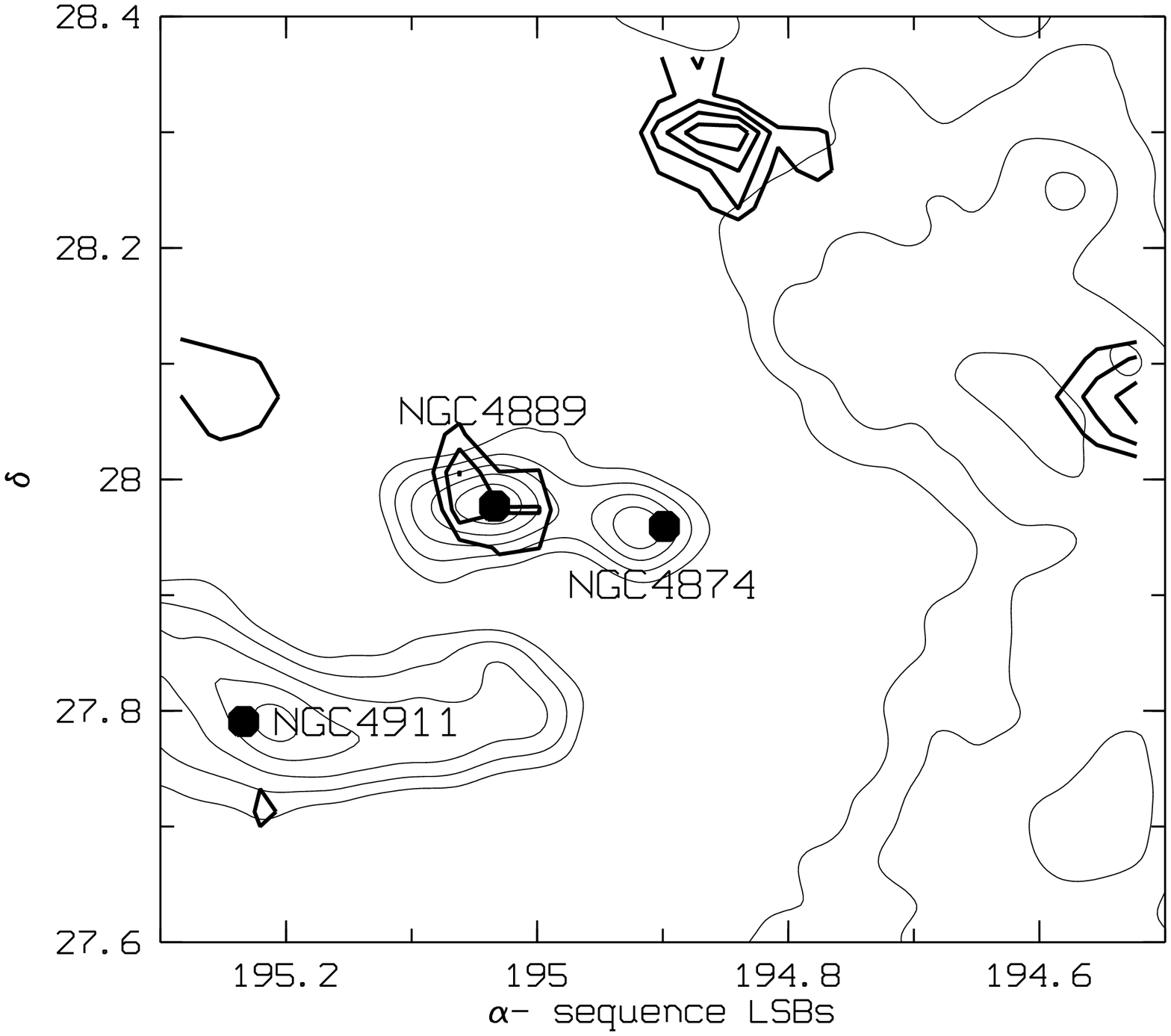,height=7.cm,angle=270}}
\caption[]{Contour plots of several fLSB samples (thick contours)
superimposed to X-ray residuals from Neumann et al. (2003) (thin
contours). Top (a): all fLSBs. Bottom (b): the CMR sequence fLSBs.
The first fLSBs contour in both plots is the 2$\sigma _d$ level and
the interval between two levels is 0.5$\sigma _d$ (see definition of
$\sigma _d$ in section 5.2).  NGC~4911, NGC~4889 and NGC~4874 are
plotted as filled dots. } \label{fig:fig16_1}
\end{figure}

\begin{figure} 
\centering
%\mbox{\psfig{figure=uplsbnew.ps,height=7.cm,angle=270}}
\mbox{\psfig{figure=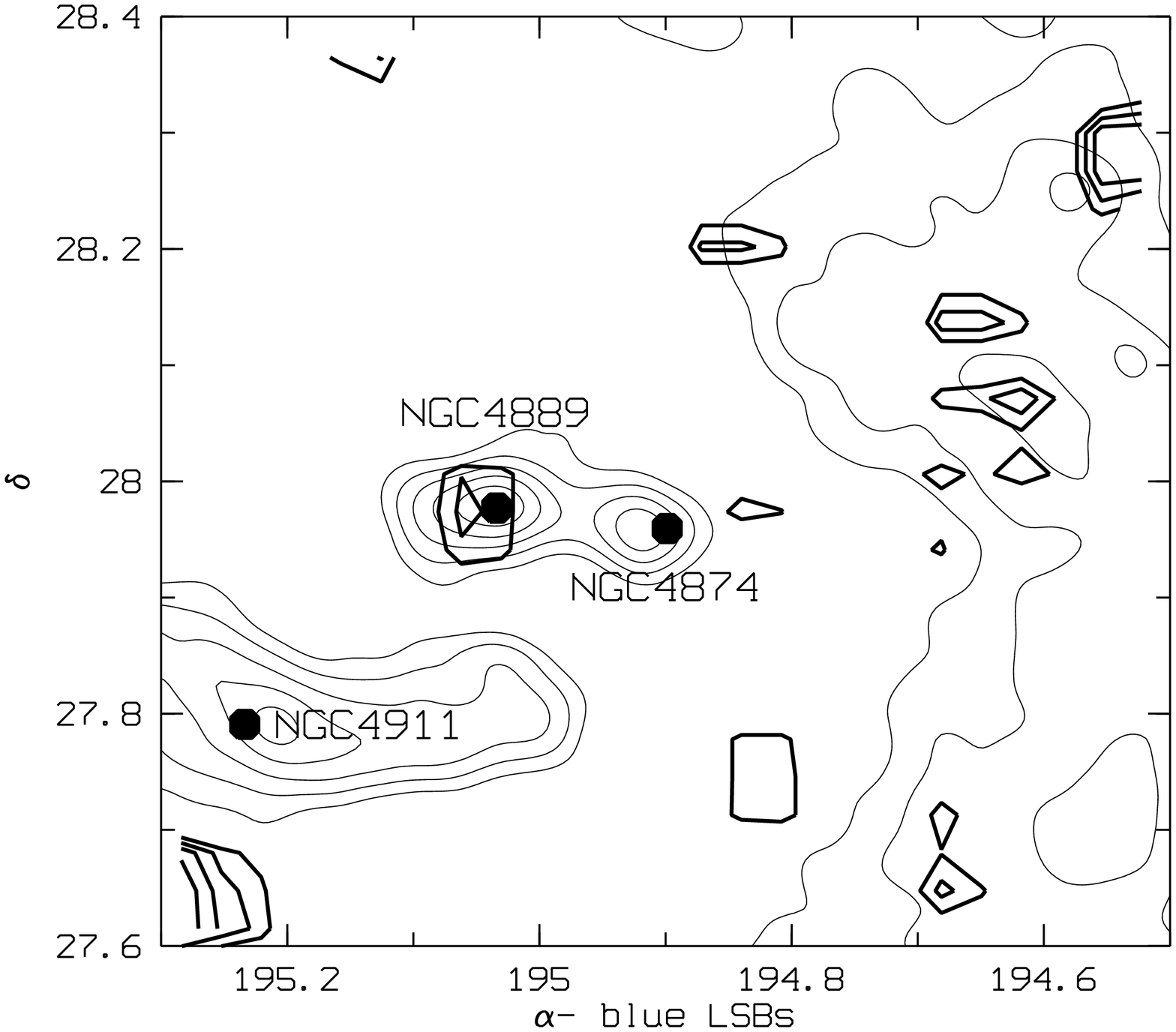,height=7.cm,angle=270}}
\caption[]{Same as Fig.~\ref{fig:fig16_1} for the blue fLSB
sample. } \label{fig:fig16_2}
\end{figure}

For discussion, we can define three regions in our color magnitude
diagram: the sequence of all the fLSBs within one standard deviation
of the main CMR red-sequence, the red population above, and the blue
population below 1 standard deviation (Fig.~\ref{fig:fig11}).

\subsection{Colors of fLSBs and empirical templates}

Empirical color templates can be used to investigate the nature of our fLSBs.
The empirical templates of Coleman et al. (1980) predict B$-$R colors
(in the exact filters we use) of 1.62 for elliptical galaxies, 1.11
for Sbc galaxies, 0.94 for Scd galaxies and 0.7 for irregular
galaxies. Fukugita et al. (1995) also predict similar colors, while
B$-$R colors of 0.74 and 0.83 are predicted for the two generic
starburst models of Kinney et al. (1996).

In this picture, we confirm that red-sequence fLSBs are indeed E-like
objects. Blue fLSBs are probably late type objects, and perhaps even
starburst galaxies for the bluest. Red fLSBs are too red to be
classical elliptical galaxies. 
These very red galaxies are as red as the bright Coma galaxies; this
should not be the case if they had simply undergone passive evolution,
given their low mass and therefore their low binding energy. Field
fLSBs are not red enough to explain this red
population and peculiar processes in the cluster must be considered,
and will be discussed in the following section.

\subsection{Colors of blank-field fLSBs}

Kolmogorov-Smirnov tests on the field and Coma fLSB distributions in
the R/B$-$R space shows (Fig.~\ref{fig:figrepres}) that the two
samples are statistically different at almost all magnitudes, except
at the Coma fLSB faint end (at R fainter than $\sim$23.65).

The slightly different seeings between the blank-field and the
Coma-field only have a minor influence on the computed colors of at
most 0.1 magnitude (the ratios between the B and R seeings are 1.188
in the Coma-field and 1.125 in the blank field).

We have no estimate of the empty field fLSBs distances, so a
direct comparison with the Coma fLSBs is impossible but we clearly see
that the empty field fLSBs are bluer than the Coma fLSBs at R brighter
than $\sim$23.4. These field galaxies probably have a higher
star-forming rate than in Coma.

\section{fLSB spatial distribution}

\subsection{fLSB properties vs. location in the cluster}

Mechanisms such as ram pressure, tidal stripping and harassment are
greatly increased in denser environments, and if fLSB formation
results from one or more of these mechanisms these galaxies should
display environment dependent structural properties.  We thus expect
fLSB structural properties to differ when fLSBs are located in the
field, in small galaxy groups, or in rich clusters (Roberts et
al. 2004, Sabatini et al. 2005), but also to vary within clusters
since galaxy cluster cores are much denser than their outer
regions. We first examine the possibility of a simple isotropic
relation between fLSB structure and cluster-centric distance. The Coma
center is taken here to be the faint galaxy center defined by Biviano et
al. (1996), very close to the X-ray center from Neumann et al. (2003).

The B$-$R colors, R magnitudes, central surface brightnesses and radii
of our fLSBs do not show significant dependence with distance to the
cluster center, in agreement with the result of Sabatini et
al. (2005) for fLSB colors in Virgo.

Besides, we can gain further details on how the cluster environment
affects fLSB formation by examining the spatial distribution of fLSBs
in the cluster. Here we compare our fLSB spatial distribution with the
X-ray residuals from Neumann et al. (2003), which indicate potential
substructures or in-falling groups. In order to investigate how fLSBs
associate with bright galaxies within Coma, we binned our fLSBs in a
two dimensional histogram with bin sizes of 1.9 arc min along right
ascension and 2 arc min along declination. We corrected the counts
using the detection efficiencies given in A06. We note that the
detection level only varies weakly from east to west across the north
and south fields, but changes strongly between the north and south
fields.

Then, in order to estimate which fLSB over densities are statistically
significant, we divided the fLSB counts by $\sigma _{distribution}$
(hereafter $\sigma _d$). $\sigma _d$ is the standard deviation of the
total number of fLSBs in the region
$\alpha$=[194.9$^\circ$,195.1$^\circ$];$\delta$=[27.6$^\circ$,27.8$^\circ$],
an area free from any peculiar fLSB density peak. The resulting fLSB
distribution is shown in Figs.~\ref{fig:fig16_1} and
~\ref{fig:fig16_2} for all fLSBs, the CMR sequence and the blue
fLSBs. The red fLSBs do not show very significant
density peaks.

\subsection{Contour plots of the whole fLSB sample}

The fLSBs from the whole sample are distributed all over the cluster
with several significant peaks (see Fig.~\ref{fig:fig16_1}a). In
particular, we detect:

- a large over-density of fLSBs around of NGC~ 4889 with a
  south extension

- a possible over-density $\sim0.22 ^\circ$ north-east of
  NGC~4889, at the border of the field

- an over-density $\sim 0.1 ^\circ$ south of NGC~4911

- an over-density $\sim 0.1 ^\circ$ west of NGC~4874

- a strong peak north of NGC~4874 at the border of the
  field. This peak does
  not appear to be associated with a specific bright
  galaxy; it could be overestimated, because the fLSB
  detection level in that region is quite low

- several peaks coinciding with the west X-ray over density

The fLSB spatial distribution suggests that  some of the Coma fLSBs 
are associated with the western X-ray over densities.  There are
over-densities near NGC~4889 significant at the 4.5$\sigma _d$ level but there
are no significant over-densities at better than the 2$\sigma _d$ level close
to NGC~4874.  This does not mean that there are $no$ fLSBs around NGC~4874, but
that their density compared to the whole field is not significantly higher. 
This suggests that the fLSBs are not directly associated with both giant 
ellipticals but only with NGC~4889.

The fLSB spatial distribution appears to be anti-correlated with that
of large scale diffuse light sources (as defined in Adami et al.
2005a). There are many fLSBs around NGC~4889 but no large scale
diffuse light structures, and few fLSBs around NGC~4874 but several
large scale diffuse light structures. The diffuse light as defined
here is so faint, that the diffuse emission does not explain the lack
of fLSBs due to detection efficiency (which has been corrected for,
but could in principle still be a problem).  This suggests that
diffuse light may have formed from the disruption of fLSBs (see also
L\'opez-Cruz et al. 1997). This hypothesis would agree with the idea
that NGC~4874 is the ``oldest'' giant galaxy in Coma {\bf (in the sense
that the dominant galaxy NGC~4874 was probably present in the Coma
cluster before NGC~4889, cf. Adami et al 2005b)}, and therefore fLSBs
neighboring NGC~4874 would have had more time to be disrupted by tidal
effects and create diffuse light (e.g. Thompson $\&$ Gregory 1993,
Lobo et al.  1997).

\subsection{Contour plots of several subsamples}

We now compare the spatial distribution of various subsamples of
fLSBs. Figs.~\ref{fig:fig16_1} and \ref{fig:fig16_2} show the
distribution of fLSBs in the CMR sequence and for blue objects.  They
exhibit features similar to the full fLSB sample with a few noticeable
differences:

- the western groups that show up when we include all fLSBs are
generally not present if we only use CMR red-sequence fLSBs, which
after detection efficiency corrections, represent 65\% of the sample

- over-densities are more prominent in the blue fLSBs which represent
22\% of the sample (about twice the red fLSBs); blue fLSBs are
distributed throughout the cluster with over-densities near the
dominant galaxies, at the field borders, but mainly in the western
X-ray extension

- the red fLSB main over-densities are located away from the center
and are not statistically significant ($\geq 2 \sigma$); however, some
of them seem correlated with the western X-ray over densities. After
detection efficiency corrections, the red fLSBs represent 13\% of the
sample.

\section{Discussion} 

For clarity, let us first summarize our main results:

i) We found 735 faint low surface brightness galaxies in the
direction of the Coma cluster, consistent with an exponential
profile when the PSF is taken into account.  The central surface
brightnesses range from $\mu _{\rm B}=24$ to 27 mag arcsec$^{-2}$,
and the total absolute magnitudes from M$_{\rm B}=-12.9$ to
M$_{\rm B}=-8.9$.

ii) From an empty field comparison, we showed that most of our 735
fLSBs are most probably members of the Coma cluster.

iii) Two thirds of fLSBs are consistent with the previously reported CMR
red-sequence for bright (R$\leq$18) ellipticals in Coma.

iv) We found a statistically significant over density of fLSBs in the
core of the cluster around NGC~4889.

v) The fLSBs exhibit no isotropic cluster-centric variations of color,
magnitude or central surface brightness, whether we consider them as a
whole or red and blue populations separately. However, blue
populations (and red ones at a lower level) are preferentially located
west of the cluster, coincident with a large X-ray over density.

We stress that this is the first large scale study of objects so faint
and difficult to detect in Coma. Recent work as e.g. Andreon \&
Cuillandre (2002), Beijersbergen et al. (2002), Iglesias-P\'aramo et
al.  (2003), or Lobo et al. (1997) did not use sufficiently deep data
to sample efficiently the regime we are studying here. Other work as
Trentham (1998) or Bernstein et al. (1995) have too small a spectral
coverage or sampled area. In the sections below we discuss our
results in terms of several possible origins of the Coma fLSBs.

\subsection{The CMR and fLSB evolution}

Along the results of Secker et al.  (1997), Odell et al. (2002)
and L\'opez-Cruz et al. (2004), we found that the CMR  in Coma can be
traced similarly for giant ellipticals and
low luminosity galaxies represented here by fLSBs. The CMR red-sequence
is described by a simple straight line fit down to total magnitudes of
 R=24.5. This effect is extremely interesting because it suggests
that galaxies over a range of more than 10 magnitudes have shared a
similar evolution process (note that Sandage 1972 had reported a
comparable spread over $\sim$8 magnitudes in Virgo).

The existence of a colour-magnitude relation for cluster galaxies implies
 that these galaxies
are made from uniformly old stellar populations (e.g. Terlevich et al.
2001). Because the CMR is a metallicity effect (e.g. Kodama \& Arimoto 1997),
elliptical galaxies experienced an extended
period of very efficient star formation at high redshift (z$\geq$2) during which
the CMR red-sequence was established by successive
generations of stars and has been evolving passively since then. Hence the
sequence fLSBs in our sample have experienced their major starburst at
the same time as the bright ellipticals and have evolved passively
since.  If we assume that sequence fLSBs are dE galaxies, as shown with color templates, this is
consistent with the results of Graham \& Guzm\'an (2003) who found a
continuous structural relation between dE and E galaxy classes in
Coma.

The formation of red-sequence 
fLSBs can be explained by a simple collapse/feedback mechanism (Dekel
\& Silk 1986).  Gas, embedded in dark matter halos present in the
universe, collapses under its own gravity to create the first
generation of stars. Because the galaxies created in this manner have
small masses and low binding potentials, the winds created by the
first supernova explosions would eject all the remaining gas out of
the galaxy and the metals they produced. This process stops star
formation and creates passively evolving fLSBs.

\subsection{Accretion of sequence fLSBs onto the Coma cluster?}

In the scenario described in the previous section, sequence fLSBs
had a formation and evolution similar to the bright cluster
galaxies. From the results of Biviano et al. (1996) and Neumann et
al. (2003) on the structure of the Coma cluster, we conclude that
these fLSBs were formed in smaller galaxy groups (along with the
bright galaxies), which later merged to create the Coma cluster as
we see it today. However, the CMR sequence fLSBs are not directly
clustered around the two dominant galaxies, but only around
NGC~4889, while the bright galaxies
are still clustered around each dominant galaxy. We can account
for this effect by evoking dynamical evolution.

Dynamical friction and relaxation, which only weakly affect low mass
galaxies, probably caused the original galaxy groups centered on
NGC~4889 and NGC~4874 to evolve in a core-halo structure with the
brightest galaxies in their core and the faintest in the halo. The two
groups then merged together.  Upon accretion onto the Coma cluster,
the faint galaxies, which generally have larger velocity dispersions
than bright galaxies, were more easily stripped off from the original
galaxy group than the bright galaxies in the tightly bound cores. The
stripped galaxies were bound to the cluster but were scattered
throughout the cluster. If NGC~4874 has been in place at the center of
the Coma cluster for a longer time than NGC~4889 (e.g. Neumann et
al. 2003), this could explain why the bright galaxies are still
clustered around the two giants but the CMR sequence fLSBs are only
clustered around NGC~4889. Besides having more time to be stripped,
the fLSBs originally around NGC~4874 could have also had more time to be
destroyed than those around NGC~4889 (e.g. Thompson $\&$ Gregory
1993). This is consistent with the suggestion by Schombert (1992) that
NGC 4874 is a cD galaxy while NGC~4889 is not.

Another scenario is to consider the central region of Coma as the merger
of a cluster containing a cD galaxy and a cluster without a cD, each
component then keeping some of its original properties.

\subsection{fLSB to giant ratio}

The fLSBs, however, cannot all be born in groups  later
accreted into clusters along with the massive galaxies due to the high
dwarf to giant ratio found in clusters (e.g.  Binggeli et al.  1990,
Sabatini et al.  2005). Sabatini et al. (2005 and references therein)
have determined that the dwarf to giant galaxy surface density ratio
is about 20 in Virgo and 4 in the Local Group.  They calculated this
ratio by simply dividing the number of galaxies brighter than M$_{\rm
B} < -19$ and those in the range M$_{\rm B}=[-14,-10$].  Based on this
result, they concluded that the dwarfs in Virgo cannot simply have
formed (via a standard CDM hierarchical scenario) around giants in the
field that fell into the potential well of Virgo. 
Some dwarf galaxies must have formed in the Virgo cluster. Similarly, Conselice et
al. (2003) show that all low mass galaxies in the Perseus cluster
cannot originate from simple early collapse.  Moore et al.
(1998) also suggested there should 
be an enhancement of dwarfs in clusters with the exception of the very
central regions where these galaxies could be destroyed. 

To compare with previous work, we computed the fLSB to giant
surface density ratio in our data. Because fLSBs are only a subsample
of dwarf galaxies, this ratio is an underestimate of the real dwarf to
giant ratio.  Using the same definition as Sabatini et al. (2005) 
for Virgo, we
found 26 giant galaxies. Among these 26, 13 have a measured redshift
and all 13 belong to the Coma cluster. We will assume therefore that
all 26 giant galaxies are part of the Coma cluster. The magnitude
range in which we detected fLSBs is M$_{\rm B}=[-12.89,-8.89]$. In
order to compare our results with Sabatini et al. (2005), we limited
the faint end to $-10$ and corrected the number of fLSBs by 7\% to
take into account the different brightness limits ($-12.89$ versus
$-14$). The 7\% value was estimated using the luminosity function of
Bernstein et al. (1995): the galaxies in the magnitude range M$_{\rm
B}=[-12.89,-10]$ account for 93\% of the galaxies in the range
M$_{\rm B}=[-14,-10]$.  The resulting number of fLSBs we would have
observed in M$_{\rm B}=[-14,-10]$ is therefore 728. We then find a
ratio of 28 to 1 for all the 728 fLSBs compared to giants. If we assume
all are in Coma, the numbers for the fLSB to giant galaxy ratio are
comparable to those in Virgo.

It is therefore tempting to conclude, as Sabatini et al. (2005) did
for Virgo, that not all the fLSBs in Coma formed around giant galaxies
prior to in fall on the Coma cluster. Additional processes, as
for example containment of metals in a giant galaxy halo by the
intracluster medium, or the formation from the remnants of stripped
galaxies are required to explain the higher fLSBs to galaxy ratio.

As shown in Fig.~\ref{fig:fig11}, the fLSBs in the blue and red
        regions may have undergone significantly different evolution
        processes from the CMR fLSBs.  This is related to the fact
        that the intrinsic fLSB B$-$R color scatter was clearly larger
        than for bright galaxies (see Section 4.1). If all fLSBs had
        followed the same evolutionary path, they should all have the
        same colors (within measurement uncertainties). The large
        color scatter is probably produced through multiple formation
        scenarios proposed below.

\subsection{Galaxy harassment and tidal stripping}

The excess of fLSBs in the cluster and their large color scatter can be partly explained if
fLSBs are the remnants of normal galaxies, which were transformed into low
        luminosity objects as they fell into the cluster.  Two such
        scenarios are ``galaxy harassment" (Moore et al. 1996) and tidal stripping.
        Galaxy harassment is defined as frequent high speed galaxy encounters which drive
        morphological transformations  as spiral
        galaxies move on their orbits across the cluster (Gallagher et al.  2001).
        Late type galaxies are disturbed by the impulse forces generated by
        these encounters that strip off mass from the galaxy, drive
        starbursts and initiate a rapid morphological evolution from large 
spirals to dwarf ellipticals (Moore et al. 1996). In contrast, tidally
stripped galaxies, which can be both spirals and ellipticals, lose stars but
do not undergo major morphological evolution. Large ellipticals are simply
transformed into dwarf ellipticals.
 Below we discuss how these two scenarios relate to red and blue fLSB formation.

Blue fLSB could be created from a re-assembly of
the outer portions of harassed spirals. One prediction of the galaxy 
harassment model is that dwarf galaxies should be assembled from the 
debris tails of the harassed galaxies (Moore et al. 1996, see also 
Barnes $\&$ Hernquist 1992, Elmegreen et al. 1993, Bournaud et al. 
2003, Duc et al. 2004 for galaxy formation scenarios in tidal tails). 
This process of dwarf galaxy formation would take place in in-falling, 
spiral-rich groups. The resulting galaxies would be bluer than 
normal, since built from external spiral parts (that are
        star forming regions) and significantly fainter. 
 Our blue fLSBs are mostly found along the line of sight of
        the large X-ray over density west of the cluster.  This X-ray 
over density is probably a collection of galaxy groups in the process 
of falling into Coma  (e.g. Neumann et al. 2003 and references therein) 
as suggested by the diffuse
        radio emission in Coma (Giovannini \& Feretti 2002). Because 
the location of blue fLSBs coincides with infalling galaxy groups, the 
blue fLSBs are likely to have been created from debris of harassed spiral galaxies.

In order to check if this scenario is viable, we compared our
results with simulations of tidal dwarf galaxy formation. Such
galaxies have masses between 10$^7$ and 10$^8$M$_\odot$ (e.g.
Bournaud et al. 2003). We estimated the masses of our fLSBs using the
M/L ratio of Mateo (1998) translated into the R band, and found that
our fLSBs have values of M/L between 5 and 20 (5 for the brightest, 20
for the faintest). This corresponds to masses between 3$\times$10$^7$
and 10$^8$M$_\odot$, typically in the range predicted by
simulations. This shows that the formation of blue  fLSBs can
occur following the scenario proposed above. 

        Red fLSBs could be the central remnants of stripped low mass early
        type galaxies.
        The metallicity is not uniformly distributed in a
        galaxy (e.g.  Zaritsky et al. 1994). There is a factor between
        3 and 5 in metallicity between the central and external parts
        of elliptical galaxies (Henry $\&$ Worthey 1999). In the
        tidal stripping scenario, fLSBs would be formed from central metal
        rich material and be redder than passively evolving galaxies
        of similar mass.  Note that the original elliptical
        galaxies would have to be already relatively faint in order to
        create red fLSBs as faint as R=21.  A similar process was also proposed
         to explain the creation of red
        low-mass galaxies in the Perseus cluster (Conselice 2002). 

        The galaxy harassment and tidal stripping scenarios are in good agreement
 with the fact
that we found the blue (and possibly the red) fLSBs to be
correlated with possible infalling groups (including spirals and
moderately early type galaxies). These infalling galaxies could have been the
source of material used to form the red and blue fLSBs.

\section{Conclusions}

Using a large sample of fLSBs detected along the Coma cluster line
of sight, we were able to reach several conclusions regarding
their various natures and their origins:

\noindent i) From the comparison with an empty field, about 95\%
of our detected fLSBs are likely to be part of the Coma cluster.

\noindent ii) Two-thirds of the fLSBs (the ones along the CMR
 sequence) experienced an evolution similar to that of bright
 ellipticals: they were formed in the same subgroups as the bright
 galaxies and joined the cluster when the subgroups merged onto
 Coma. Upon accretion, some fLSBs were ejected from the subgroups,
 scattered throughout the cluster, but yet retained by the underlying
 cluster potential. They have been undergoing passive evolution since
 then. Some could also have been destroyed following the Thompson $\&$
 Gregory (1993) scenario. If passive evolution is the explanation for
 the red-sequence fLSBs, the fact that the ratio of red-sequence fLSBs
 to giant galaxies in Coma is still significantly higher than in the
 field remains a puzzle. Only numerous fusions of bright galaxies
 could explain this ratio. We plan to investigate this question in a
 future paper.

\noindent iii) fLSBs that fall on the color magnitude relation are
consistent with a simple collapse feedback scenario: fLSBs were
formed when gas collapsed and ignited a starburst. Because of
their small size and low binding potential, supernova winds could
have ejected all the remaining gas and metals from the galaxy and
halted additional star formation.

\noindent iv) We found indications supporting a scenario in
which debris from galaxy harassment can create blue fLSBs.

\noindent v) Formation of red fLSBs as the central remnants of
small stripped early type galaxies is also a possibility.

Further studies involving redshifts,
        velocities  and gas  content measurements  are needed  to gain
        more clues  on the formation and evolution  processes of these
        fLSBs.

\begin{acknowledgements}
The authors thank the referees for useful and constructive
comments and are grateful to the CFHT and Terapix teams, and to
the French CNRS/PNG for financial support. Some of the authors
also acknowledge support from NASA Illinois space grant
NGT5-40073, from
Northwestern University and from NSF grant AST-0205960.\\
\end{acknowledgements}

\end{document}